\documentclass[sigconf]{acmart}
\acmSubmissionID{698}

\usepackage{amsmath}               
  {
      \theoremstyle{plain}

  }
\usepackage{enumitem}
\AtBeginDocument{%
  \providecommand\BibTeX{{%
    \normalfont B\kern-0.5em{\scshape i\kern-0.25em b}\kern-0.8em\TeX}}}

\usepackage{tikz}
\usepackage{subcaption}
\usepackage{pgfplots}
\usetikzlibrary{arrows,positioning,automata,calc,shapes}
\pgfplotsset{compat=newest, scaled z ticks=false} 
\pgfplotsset{plot coordinates/math parser=false}
\newlength\figureheight 
\newlength\figurewidth
\usepackage{tikz}
\usepackage{caption}
\usepackage{graphicx}
\usepackage{subcaption}
\usepackage{filecontents}
\usepackage{float}
\usepackage{url}
\usepackage{amsmath}
\usepackage{graphicx}
\usepackage{url}
\usepackage{setspace}
\usepackage{hyperref}

\usepackage{amsfonts,amssymb}
\usepackage{color, soul}
\usepackage{mathrsfs}
\usepackage{cleveref}
\usepackage{multirow}
\usepackage{wrapfig}
\usepackage{amsthm}
\usepackage{algorithm}
\usepackage{algorithmic}
\usepackage{bm}
\usepackage{mdframed}
\usepackage{supertabular}

\usepackage{balance}
\definecolor{lightcrimson}{rgb}{0.93, 0.16, 0.51}

\AtBeginDocument{%
  \providecommand\BibTeX{{%
    Bib\TeX}}}

\copyrightyear{2025}
\acmYear{2025}
\setcopyright{acmlicensed}\acmConference[WWW '25]{Proceedings of the ACM Web Conference 2025}{April 28--May 2, 2025}{Sydney, Australia}
\acmBooktitle{Proceedings of the ACM Web Conference 2025 (WWW '25), April 28--May 2, 2025, Sydney, Australia}

\author{Yaochen Zhu}
\authornote{Work done when Yaochen and Chao were machine learning research interns at Netflix.}
\email{uqp4qh@virginia.edu}
\affiliation{%
\institution{University of Virginia}
\city{Charlottesville}
\state{VA}
\country{USA}
\country{}
}

\author{Chao Wan}
\email{cw862@cornell.edu}
\affiliation{%
\institution{Cornell University}
\city{Ithaca}
\state{NY}
\country{USA}
}

\author{Harald Steck}
\email{hsteck@netflix.com}
\affiliation{%
\institution{Netflix Inc.}
\city{Los Gatos}
\state{CA}
\country{USA}
}

\author{Dawen Liang}
\email{dliang@netflix.com}
\affiliation{%
\institution{Netflix Inc.}
\country{}
\city{Los Gatos}
\state{CA}
\country{USA}
}

\author{Yesu Feng}
\email{yfeng@netflix.com}
\affiliation{%
\institution{Netflix Inc.}
\country{}
\city{Los Gatos}
\state{CA}
\country{USA}
}

\author{Nathan Kallus}
\email{nkallus@netflix.com}
\affiliation{%
\institution{Netflix Inc. \& Cornell University}
\country{}
\city{New York}
\state{NY}
\country{USA}
}
\author{Jundong Li}
\email{jundong@virginia.edu}
\affiliation{%
\institution{University of Virginia}
\city{Charlottesville}
\state{VA}
\country{USA}
}

\renewcommand{\shortauthors}{Yaochen Zhu et al.}



\copyrightyear{2025}
\acmYear{2025}
\setcopyright{cc}
\setcctype{by-nc-sa}
\acmConference[WWW '25]{Proceedings of the ACM Web Conference 2025}{April 28-May 2, 2025}{Sydney, NSW, Australia}
\acmBooktitle{Proceedings of the ACM Web Conference 2025 (WWW '25), April 28-May 2, 2025, Sydney, NSW, Australia}
\acmDOI{10.1145/3696410.3714908}
\acmISBN{979-8-4007-1274-6/25/04}

\begin{document}

\title{Collaborative Retrieval for Large Language Model-based Conversational Recommender Systems}

\begin{abstract}

\noindent Conversational recommender systems (CRS) aim to provide personalized recommendations via interactive dialogues with users. While large language models (LLMs) enhance CRS with their superior understanding of context-aware user preferences, they typically struggle to leverage behavioral data, which have proven to be important for classical collaborative filtering (CF)-based approaches. For this reason, we propose \texttt{CRAG}—\underline{\textbf{C}}ollaborative \underline{\textbf{R}}etrieval \underline{\textbf{A}}ugmented \underline{\textbf{G}}eneration for LLM-based CRS. To the best of our knowledge, \texttt{CRAG} is the first approach that combines state-of-the-art LLMs with CF for conversational recommendations. Our experiments on two publicly available movie conversational recommendation datasets, i.e., a refined Reddit dataset (which we name Reddit-v2) as well as the Redial dataset, demonstrate the superior item coverage and recommendation performance of \texttt{CRAG}, compared to several CRS baselines. Moreover, we observe that the improvements are mainly due to better recommendation accuracy on recently released movies. The code and data are available at {\color{lightcrimson}\url{https://github.com/yaochenzhu/CRAG}.}

\vspace{-2mm}

\end{abstract}

\maketitle

\section{Introduction}

Recommender systems (RS) have become an indispensable component on digital service platforms  \citep{jannach2010recommender}. Traditional RSs, such as collaborative filtering \cite{koren2021advances}, have demonstrated effectiveness in leveraging historical user-item interactions for recommendations. Conversational recommender systems (CRS) create a more engaging and interactive environment for users---they enable users to express preferences freely in natural language and refine their thoughts through multiple rounds of interactions \citep{sun2018conversational,jannach2021survey}, where more precise and personalized recommendations can be offered to users.

\begin{figure}[t]
\centering
\includegraphics[width=0.7\linewidth]{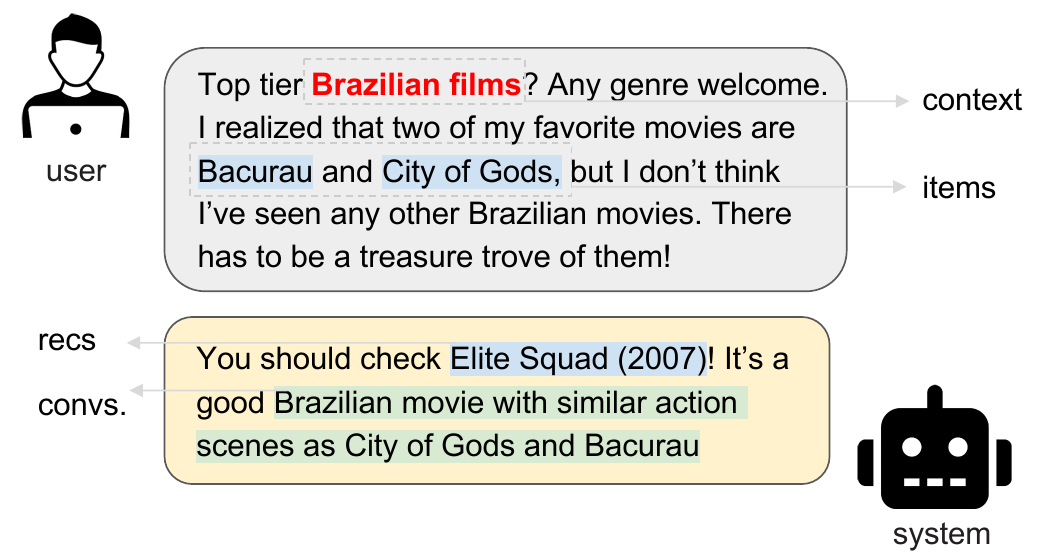}
\vspace{-3mm}
\caption{An example for conversational recommendations, with items and relevant context highlighted in user query.}
\vspace{-5mm}
\label{fig:teaser}
\end{figure}

Compared with traditional RSs, CRSs need to comprehensively consider both \textit{items} and \textit{context} in the dialogue, which is essential for user preference understanding and recommendation generations (see Fig. \ref{fig:teaser}). Early CRSs \cite{li2018towards, sun2018conversational} used traditional RS models, such as factorization machines \cite{rendle2010factorization}, and sequential models, such as recurrent neural networks (RNN) \cite{chung2014empirical}, to separately model items and context in the dialogue, where external item/word knowledge graphs (e.g., DBpedia \cite{auer2007dbpedia} and ConceptNet \cite{speer2017conceptnet}) are often leveraged to provide additional information. Subsequently, transformers \cite{vaswani2017attention} pretrained on external corpora (e.g., GPT-2 \cite{radford2019language}) were introduced to enrich item/context representations with prior knowledge \cite{zhou2020improving,wang2022towards,feng2023large}. Meanwhile, semantic fusion strategies such as cross-attention \cite{chen2019towards,zhou2020improving}, mutual information maximization \cite{wang2022towards}, and contrastive learning \cite{zhou2022c2} were developed to integrate the representations of items and context to model context-aware user preferences. 

Recently, large language models (LLMs), such as GPT-4o \cite{openai2024} and Claude 3.5-Sonnet \cite{anthropic2024}, have demonstrated an unprecedented understanding of both items and context in natural language \cite{zhao2023survey}. Pretrained on vast corpora across various domains, these LLMs can be viewed as unstructured knowledge databases that encompass extensive knowledge of items and their relations \cite{ren2024representation}. For instance, \citet{xi2023towards} showed that item knowledge prompted out from LLMs can enhance the recommendation accuracy of traditional RSs. Additionally, with their strong reasoning abilities, LLMs can generate more accurate recommendations by deriving user preferences based on better considerations of both items and context in the dialogue \cite{wei2024llmrec}. Built upon the recent advances in LLMs, \citet{he2023large} demonstrated that these powerful models (e.g., GPT-4o) can serve as good zero-shot CRSs, which substantially improve the recommendation performance compared with traditional CRS methods even though they are not directly trained on the CRS data. 

While state-of-the-art LLMs possess extensive knowledge and reasoning abilities, they typically fall short in leveraging collaborative filtering (CF), a fundamental and effective technique in traditional RS \cite{zhu2024collaborative,zheng2024adapting,wu2024coral}. The reason is that user-item interaction data are usually proprietary (and therefore not included in the LLMs' training corpora) and difficult to be fully described in natural language. Moreover, even if CF information can be integrated into LLM-based CRS, existing research indicates that adding more external knowledge does not necessarily enhance the LLMs (which are already very powerful) \cite{gao2023retrieval}, as it can introduce noise that biases their behavior. Therefore, effectively utilizing CF information to complement the \textit{context} in the dialogue and LLMs' inherent \textit{content} knowledge presents a significant challenge for LLM-based CRS. As an aside, in a different line of work, CF has been used for improving LLMs in the classical recommendation setting \citep{wu2024survey}: Most works focus on \textit{white-box} LLMs, where the model weights are accessible to the researcher \citep{zhu2024collaborative,bao2023tallrec,hua2023index,kim2024large, zheng2024adapting}.
 White-box LLMs are generally smaller in scale compared to large proprietary LLMs, which are typically much more powerful, both in terms of their knowledge and reasoning capabilities. Due to the inaccessibility of model weights, however, combining CF with \textit{black-box} LLMs is comparatively less explored  \citep{xi2024towards,wu2024coral}
(see Appendix \ref{sec:rel_work} for more detailed discussions).

In this paper, to improve upon zero-shot LLMs, i.e., the current state-of-the-art \citep{he2023large},
 we propose \texttt{CRAG}, i.e., \underline{\textbf{C}}ollaborative \underline{\textbf{R}}etrieval \underline{\textbf{A}}ugmented \underline{\textbf{G}}eneration for LLM-based CRSs.
To the best of our knowledge, \texttt{CRAG} is the first approach that combines state-of-the-art, black-box LLMs with collaborative filtering in the scenario of \textit{conversational recommendations}, where context-aware CF knowledge can be introduced to enhance the recommendation performance. In our experiments in Sections \ref{sec:exp} and \ref{sec:exp2}, we show that \texttt{CRAG} leads to improved recommendation accuracy on two publicly available movie conversational recommendation datasets. We also provide several ablation studies to shed light on the inner workings of \texttt{CRAG} in Sections \ref{sec:abl} and \ref{sec:abl2}. Apart from that, we establish and release a refined version of the Reddit dataset \cite{he2023large} on movie recommendations, where the extraction of movies mentioned in the dialogues is substantially improved (see Section \ref{sec:data}). We also show (see \textbf{Finding 1} in Section \ref{sec:data}) that this improvement in extraction accuracy can have a considerable impact on the derived insights. 

\section{Problem Formulation}

In this section, we formally define the CRS problem that we study in this paper. Let $\mathcal{U}$ denote the set of users and $\mathcal{I}$ the set of items. A conversation between a user and the CRS is denoted as $C = \{(u_t, s_t, \mathcal{I}_{t})\}_{t=1}^T$, where at the $t$-th turn, $u_t \in \{\texttt{User}, \texttt{System}\}$ generates an utterance $s_t = (w_1, w_2, \dots, w_{N_t})$, which is composed of $N_t$ tokens from the vocabulary $\mathcal{V}$. $\mathcal{I}_{t}$ denotes the set of items mentioned in $s_t$. We assume that users can freely mention any item from $\mathcal{I}$ in the query, but the system can only recommend items from a fixed catalog (e.g., available movies on a specific platform like Netflix). We use $\mathcal{Q} \subseteq \mathcal{I}$ to denote the catalog of items available for recommendations. Here, we note that $\mathcal{I}_{t}$ is usually \textbf{not} annotated by the user and may be empty if no items are mentioned at the $t$-th turn. The CRS backbone is a black box LLM $\Phi$, with historical interaction data $\mathbf{R} = \{0, 1\}^{|\mathcal{U}_{r}| \times |\mathcal{I}|}$ available as an external collaborative filtering knowledge database. Users in $\mathcal{U}_{r}$ do not have to be the same as $\mathcal{U}$. $\mathbf{r}_{u \in \mathcal{U}_{r}} \in \{0, 1\}^{|\mathcal{I}|}$ denotes the behavior patterns of user $u$, and is generally not included in the LLM training corpora.

The focus of this paper is mainly on the recommendation part of CRS, which aims to generate a ranked list of items $\hat{\mathcal{I}}_{k}$ from the catalog $\mathcal{Q}$ based on the current dialogue $C_{:k-1} = \{(u_t, s_t, \mathcal{I}_{t})\}_{t=1}^{k-1}$ and the available interaction data $\mathbf{R}$, such that the generated $\hat{\mathcal{I}}_{k}$ best matches the groundtruth items in $\mathcal{I}_{k}$ (if $\mathcal{I}_{k} \neq \emptyset$ and $u_{k} = \texttt{System}$).

\begin{figure*}[t]
\centering
\includegraphics[width=0.8\linewidth]{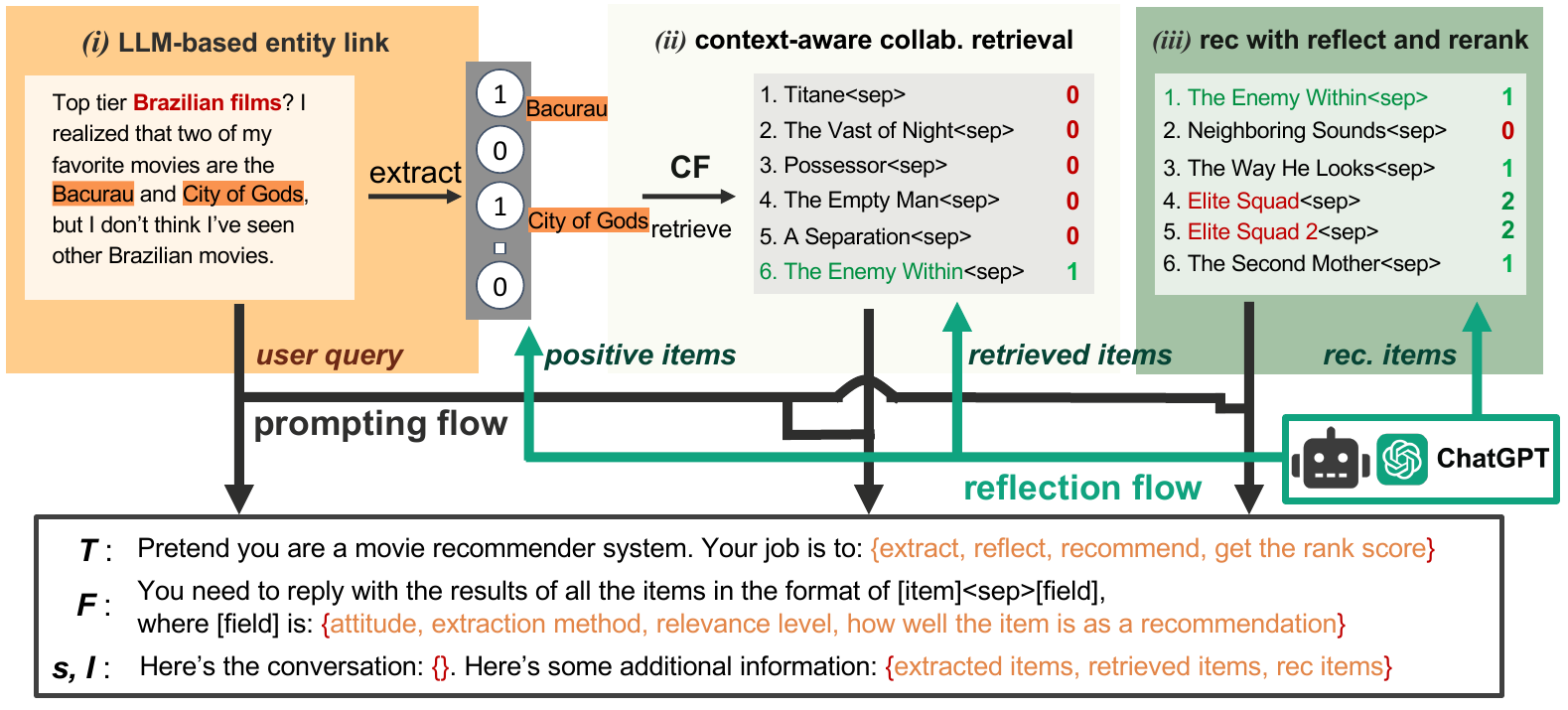}
\caption{Overview of CRAG for CRS and its three components: \textbf{\textit{(i)}} LLM-based entity link; \textbf{\textit{(ii)}} context-aware collaborative retrieval, and \textbf{\textit{(iii)}} recommendation with reflect and rerank. The reflection steps are emphasized in green arrows. The sub- and super-script for different task-specific prompt $T$, format instruction $F$, and item list $\mathcal{I}$ are omitted for simplicity.}
\label{fig:framework}
\end{figure*}

\section{Approach}\label{sec:approach}
In this section, we introduce \texttt{CRAG}, a collaborative retrieval-aug-mented LLM for conversational recommendations. The overall framework of \texttt{CRAG} is illustrated in Fig. \ref{fig:framework}. \texttt{CRAG} consists of three components,  \textbf{\textit{(i)}} \textit{LLM-based entity link}, \textbf{\textit{(ii)}} \textit{collaborative retrieval with context-aware reflection},  
 and \textbf{\textit{(iii)}} \textit{recommendation with reflect-and-rerank}, which will be outlined in the following parts.

\subsection{LLM-based Entity Link}
\label{sec:entity}

\textit{Entity linking}, i.e., extracting items $\mathcal{I}_{k}$ from each utterance $s_{k}$ and mapping them to the item database $\mathcal{I}$, is crucial for most CRS frameworks, as it bridges the gap between textual dialogues and external structured knowledge (e.g., knowledge graphs for traditional CRSs and interaction data $\mathbf{R}$ for \texttt{CRAG}). However, existing methods, e.g., Bayesian models \cite{daiber2013improving} or supervised finetuning of transformers \cite{he2023large}, rely on simulated data with seed items and struggle with handling abbreviations, typos, and ambiguity in item titles. Consequently, entity recognition noise is pervasive for the current CRS methods. 

\subsubsection{\textbf{LLM-based Entity Extraction}} In \texttt{CRAG}, we leverage the pretrained knowledge and reasoning ability of LLMs to extract mentioned items in each utterance $s_{t}$. Additionally, we analyze the \textit{attitude} associated with each item to capture the sentiment or stance context under which the items are mentioned by the user in the dialogue. This process (Fig. \ref{fig:framework}-\textbf{\textit{(i)}}) can be formally denoted as:
\begin{equation}
\label{eq:ent_ext_raw}
\mathcal{I}^{raw}_{t} = f_{e}\left(\Phi\left(T_{e}, F_{e}, s_{t}\right)\right).
\end{equation}
Here, $T_{e}$ is a task-specific prompt\footnote{The details of the prompts defined in the main paper are provided in Appendix \ref{sec:prompts}.} that instructs the LLM $\Phi$ to reply with the standardized form of the items mentioned in the utterance $s_{t}$ given that potential abbreviations, typos, and ambiguity could exist in $s_{t}$. In addition, to further improve the extraction efficiency, we design a \textit{batch inference format instruction} $F_{e}$ to guide the LLM to reply with all the item-attitude pairs in the utterance $s_{t}$ in the form of \texttt{"[item]<sep>[attitude]"}, where we empirically set \texttt{<sep>} to {"\#\#\#\#"} as the dummy tokens that separate the item name and the associated attitude in the response. In $F_{e}$, we explicitly instruct the LLM to output attitudes as numerical values in the range \texttt{\{-2, -1, 0, 1, 2\}}, representing attitude categories in the spectrum of \texttt{\{very negative, negative, neutral, positive, very positive\}}. This numerical encoding helps minimize errors in the generations. With $F_{e}$, the raw set of item-attitude pairs $\mathcal{I}^{raw}_{t} = \left\{\left(i^{raw}_{t,j}, a_{t,j}\right)\right\}_{j}$ can be trivially extracted from the LLM’s output using a string processing function $f_{e}$ that parses lines and the \texttt{<sep>} tokens.  

\subsubsection{\textbf{Bi-level Match and Reflection}}

In the current stage, each raw item $i^{raw}_{t,j} \in \mathcal{I}^{raw}_{t}$ is a text string that may still be in non-standardized forms or contain small typos. To accurately link each raw item $i^{raw}_{t,j}$ to the item database $\mathcal{I}$, we introduce a bi-level match and reflection module that combines \textit{character}-level and \textit{word}-level fuzzy match with LLM-based reflection to post-fix the disagreements. Specifically, character-level match addresses typos in $i^{raw}_{t,j}$ \cite{he2023large}, whereas word-level match links certain abbreviations (e.g., \texttt{"Star Wars I"}) to their full names in the database $\mathcal{I}$ (e.g., \texttt{"Star Wars I - The Phantom Menace"}). We denote the two candidate sets produced by above match processes as $\mathcal{I}^{char}_{t}$, $\mathcal{I}^{word}_{t}$. Furthermore, we ask the LLM to reflect on the disagreements (if any) between $\mathcal{I}^{char}_{t}$ and $\mathcal{I}^{word}_{t}$, which is formally denoted as follows:
\begin{equation}
\label{eq:ent_ext_ref}
\mathcal{I}^{ref}_{t} = f^{ref}_{e}\left(\Phi \left(T^{ref}_{e}, F^{ref}_{e}, \mathcal{I}^{char}_{t}, \mathcal{I}^{word}_{t}, s_{t}\right)\right).
\end{equation}
In this step, the task-specific prompt $T^{ref}_{e}$ instructs the LLM to reflect on the differences between $\mathcal{I}^{char}_{t}$ and $\mathcal{I}^{word}_{t}$ based on the utterance $s_{t}$. In addition, the \textit{batch reflection format instruction} $F^{ref}_{e}$ guides the LLM to judge all the disagreements simultaneously and return the final reflection result of each item in the format of \texttt{"[matched\_item]<sep>[method]"}, where \texttt{"[matched\_item]"} is the item that the LLM determines to be correctly linked to the database $\mathcal{I}$ (could be empty if none is found), and \texttt{"[method]"} in \texttt{\{char, word, both, none\}} indicates the correct matching strategy. Finally, the function $f^{ref}_{e}$ processes the LLM's output by selecting, removing, or correcting each item based on the \texttt{"[matched\_item]"} and \texttt{"[method]"} fields to form the final item set $\mathcal{I}^{ref}_{t}$ for $s_{t}$.

\subsection{Context-Aware Collaborative Retrieval}

After extracting and linking items for each utterance $s_{t}$ in the dialogue $C_{:k-1}$ to the database $\mathcal{I}$, we introduce the collaborative retrieval module of \texttt{CRAG}. This module aims to retrieve context-relevant items based on the current dialogue $C_{:k-1}$ and historical interactions $\mathbf{R}$, which augments the prompt with collaborative filtering (CF) knowledge to enhance the LLM-based recommendations.

\subsubsection{\textbf{Collaborative Retrieval}} Collaborative retrieval, similar to other retrieval-augmented generation (RAG) strategies \cite{gao2023retrieval}, follows two main steps: query rewriting and similarity matching. The overall process for collaborative retrieval is defined as follows:
\begin{equation}
\label{eq:cr}
    \mathcal{I}^{CR}_{k} = \operatorname{Top}_{K}({Sim} \left(f_{r}\left(C_{:k-1}), \mathcal{Q}; \mathbf{R}\right)\right),
\end{equation}
where the query rewrite function $f_{r}(C_{:k-1})$ aggregates the positively mentioned items from the current dialogue $C_{:k-1}$, i.e., $\mathcal{I}^{q}_{k} = \cup^{k-1}_{t=1} \mathcal{I}_{t}$, and converts them into a multi-hot variable $\mathbf{r}_{k} \in \{0, 1\}^{|\mathcal{I}|}$. Since it is generally risky to extrapolate negatively mentioned items through CF (as the reason for disliking an item tends to be more subjective than collaborative) and because of the small number of negative item mentions in the dialogues (see Fig. \ref{fig:attitude} in the Appendix), we exclude these items from the collaborative retrieval model. Afterward, we retrieve the top-$K$ items from the catalog $\mathcal{Q}$ based on their collaborative similarity (measured via the $Sim$ function derived from the interaction data $\mathbf{R}$) with the items in $\mathcal{I}^{q}_{k}$.

Various CF methods \cite{mnih2007probabilistic,liang2018variational} can be used to learn the $Sim$ function based on the interaction data $\mathbf{R}$. In this paper, we utilize a simple while effective adapted EASE \cite{steck2019embarrassingly} objective as follows: 
\begin{equation}
\label{eq:cf_retrieve}
\begin{aligned}
\min _\mathbf{W} \ \ \ & \|\mathbf{R}_{\mathcal{Q}}-\mathbf{R} \mathbf{W}\|_F^2+\lambda \cdot\|\mathbf{W}\|_F^2 \\
\text { s.t. } & \mathbf{W}_{i,j}=0, \forall i=\mathrm{ReID}(j),
\end{aligned}
\end{equation}
where $\mathbf{R}_{\mathcal{Q}}$ selects the columns in $\mathbf{R}$ that correspond to the items in the catalog $\mathcal{Q}$, the asymmetric matrix $\mathbf{W} \in \mathbb{R}^{|\mathcal{I}| \times |\mathcal{Q}|}$ maps the space of items that users mention freely in the dialogue (i.e., $\mathcal{I}$) to the space of items available for recommendation in the catalog $\mathcal{Q}$, and the function $\mathrm{ReID}$ remaps the indices of the catalog items from $\mathcal{I}$ to $\mathcal{Q}$. The constraint in Eq. (\ref{eq:cf_retrieve}) prevents self-reconstruction from being used as a shortcut for the similarity matrix $\mathbf{W}$. Based on Eq. (\ref{eq:cf_retrieve}), the similarity function is then defined as $Sim(\mathcal{I}^{q}_{k}, \mathcal{Q}; \mathbf{R}) = \mathbf{r}^{T}_{k} \times \mathbf{W}$, which returns the similarity score of each item in $\mathcal{Q}$ relative to the positively mentioned items in $C_{:k-1}$. The scores are then used to select items in the collaborative retrieval $\mathcal{I}^{CR}_{k}$. In addition, $\mathbf{W}$ is adjusted by more recent item-popularities based on \cite{steck2019high}.

\subsubsection{\textbf{Context-Aware Reflection}}

Since the raw collaborative retrieval defined in Eq. (\ref{eq:cf_retrieve}) does not consider any context information in the current dialogue $C_{:k-1}$, directly augmenting the retrieved items $\mathcal{I}^{CR}_{k}$ in the prompt as extra collaborative knowledge could introduce context-irrelevant information, thereby biasing the LLM's recommendations. To address this issue, we post-process the retrieved items via an LLM-based context-aware reflection step as:
\begin{equation}
\label{eq:ref_tre}
\mathcal{I}^{aug}_{k} = f^{aug}\left(\Phi\left(T^{aug}, F^{aug}, C_{:k-1}, \mathcal{I}^{CR}_{k}\right)\right),
\end{equation}
where $T^{aug}$ is the task-specific prompt that instructs the LLM to reflect on the contextual relevancy of items in $\mathcal{I}^{CR}_{k}$ based on the dialogue $C_{:k-1}$. In addition, $F^{aug}$ is the \textit{context-relevance batch reflection instruction} that guides the LLM to reply with the simultaneous judgment of all the items in $\mathcal{I}^{CR}_{k}$ in the format of \texttt{"[item]<sep>[relevance]"}, where \texttt{[relevance]} is a binary score in \texttt{\{0, 1\}} indicating whether or not a retrieved \texttt{[item]} is contextually relevant. After the reflection, only items that are judged as context-relevant are preserved in $\mathcal{I}^{aug}_{k}$, i.e., the context-aware collaborative retrieval, which is ready to be augmented into the prompt for recommendation generations. For example, in the example illustrated in Fig. \ref{fig:framework}-\textbf{\textit{(ii)}}, although all the retrieved movies are similar to \texttt{City of God} and $\texttt{Bacurau}$, only \texttt{The Enemy Within} is Brazilian, where the rest are removed from $\mathcal{I}^{aug}_{k}$ after the reflection.

\subsection{Recommendation with Reflect and Rerank}

In this section, we discuss the \textit{generation} phase of \texttt{CRAG}, which generates the final recommendation list with LLM based on the reflected collaborative retrieval $\mathcal{I}^{aug}_{k}$. This phase consists of three key steps: \textbf{\textit{(i)}} collaborative query augmentation, \textbf{\textit{(ii)}} LLM-based item generation, and \textbf{\textit{(iii)}} reflect and rerank (post-processing).

\subsubsection{\textbf{Collaborative Query Augmentation}}

The preliminary step of utilizing the context-aware collaborative knowledge in $\mathcal{I}^{aug}_{k}$ is to augment it into the prompt for recommendations. This starts with adding a pretext to emphasize the collaborative nature of the retrieved items, such as: \texttt{"Below are items other users tend to interact with given the positive items mentioned in the dialogue:"}. Afterward, $\mathcal{I}^{aug}_{k}$ is transformed into a string, i.e., $I^{aug}_{s,k}$, that lists the similarity-ranked items separated by semicolons.

We note that $I^{aug}_{s,k}$ opens up to two interpretations in \texttt{CRAG}. From a RAG perspective, $I^{aug}_{s,k}$ serves as the extra CF information retrieved from an external user-item interaction database $\mathbf{R}$; from a recommendation perspective, $I^{aug}_{s,k}$ also represents the possible item candidates that could be used in the final recommendations. Based on these interpretations, we design two distinct prompts to instruct the LLM on how to use the augmented collaborative information: \textbf{\textit{(i)}} a \textit{rag} prompt that instructs the LLM to use the augmented information at its own discretion.
\textbf{\textit{(ii)}} a \textit{rec} prompt that explicitly asks the LLM to consider the augmented items as candidates for recommendations (see Appendix \ref{sec:prompts}). Empirically, we find that different prompts work for different models. For example,  GPT-4o enjoys the freedom in the \textit{rag} prompt, whereas GPT-4 tends to ignore the retrieved items in $I^{aug}_{s,k}$ under the same prompt and instead needs the \textit{rec} prompt to force it to consider the items in $I^{aug}_{s,k}$.

\subsubsection{\textbf{LLM-based Recommendations}}\label{sec:rec_step}

After constructing the collaborative augmentation $I^{aug}_{s,k}$ from the item list $\mathcal{I}^{aug}_{k}$, it is appended to the current dialogue $C_{:k-1}$ and input into the LLM to generate a preliminary recommendation list. The collaborative augmented generation step in \texttt{CRAG} is formalized as follows:
\begin{equation}
\label{eq:rec}
\mathcal{I}^{rec}_{k} = f^{rec}\left(\Phi\left(T^{rec}, F^{rec}, C_{:k-1}, I^{aug}_{s,k}\right)\right),
\end{equation}
where the prompt $T^{rec}$ instructs the LLM to function as a CRS that generates a ranked item list as recommendations based on both the dialogue $C_{:k-1}$ and the collaborative augmentation $I^{aug}_{s,k}$. The format instruction $F^{rec}$ guides the LLM to return the standardized item names seperated in lines. Eq. (\ref{eq:rec}) takes into account both the dialogue context and the collaborative information to generate the recommendations, thereby addressing the key limitations of zero-shot LLM-based CRS: the lack of collaborative filtering abilities.

\subsubsection{\textbf{Reflect and Rerank}}\label{sec:rerank_step}

While the context-aware collaborative knowledge in $I^{aug}_{s,k}$ substantially enhances the relevancy of generated recommendations, it can also trigger a bias inherent in LLMs, where the attention mechanism tends to replicate the retrieved items in $I^{aug}_{s,k}$ at the beginning of the recommendations. Since items in $I^{aug}_{s,k}$ are retrieved by considering only the collaborative information (as the context-aware reflection in Eq. (\ref{eq:ref_tre}) only removes items), the most relevant items in $\mathcal{I}^{rec}_{k}$ generated by LLM (which are not necessarily in $I^{aug}_{s,k}$) may not be ranked on the top.

Here, a naive approach to mitigate the bias is to directly ask the LLM to rerank the items in the raw recommendations $\mathcal{I}^{rec}_{k}$. However, this may result in a nonsensical reranked list with missing items, probably due to the large semantic gap between the input items $\mathcal{I}^{rec}_{k}$ and the asked target, i.e., reranked items in $\mathcal{I}^{rec}_{k}$ based on the context relevancy. To bridge the gap, we propose a reflect-and-rerank module that asks the LLM to assign ordinal scores to each item in $\mathcal{I}^{rec}_{k}$ based on how well it aligns as a recommendation based on the dialogue $C_{:k-1}$. This can be  formalized as:
\begin{equation}
\label{eq:ref_rerank}
\mathcal{I}^{r\&r}_{k} = f^{r\&r}\left(\Phi\left(T^{r\&r}, F^{r\&r}, C_{:k-1}, \mathcal{I}^{rec}_{k}\right)\right),
\end{equation}
where the task-specific prompt $T^{r\&r}$ instructs the LLM to reflect on the recommendations and assign scores to all the items in $\mathcal{I}^{rec}_{k}$ based on the dialogue $C_{:k-1}$. In addition, the \textit{batch reflect-and-rerank instruction} $F^{r\&r}$ guides the LLM to return the scores for all the items in $\mathcal{I}^{rec}_{k}$ simultaneously in the format \texttt{"[item]<sep>[score]"}, where \texttt{[score]} $\in$ \texttt{"\{-2, -1, 0, 1, 2\}"} corresponds to the level of recommendation quality in \texttt{\{very bad, bad, neutral, good, very good\}}. These scores serve as a reference for evaluating the relative suitability of each item, providing an intermediate step to address the semantic gap between the input item list $\mathcal{I}^{rec}_{k}$ and the context-aware reranked item list $\mathcal{I}^{r\&r}_{k}$. From the example in Fig. \ref{fig:framework}-\textbf{\textit{(iii)}}, we can see that even though \texttt{The Enemy Within} is a good recommendation based on the collaborative information, after the reflect-and-rerank step of \texttt{CRAG}, more relevant ones such as \texttt{"Elite Squad"} and \texttt{"Elite Squad 2"} can be reranked on the top.

\subsection{Conversations without Item Mentions}
\label{sec:coldstart}

Finally, to introduce the context-aware CF information to improve the recommendations when user mentions no items in the dialogue $C_{:k-1}$, we first prompt the LLM to infer potential items the user might like based on $C_{:k-1}$. The generated items are then mapped to the database $\mathcal{I}$ via the strategy in Section \ref{sec:entity}, which can be treated as $\mathcal{I}^{q}_{k}$ in Eq. (\ref{eq:cr}), and the remaining parts of \texttt{CRAG} remain the same.

\section{Empirical Study}

\subsection{CRS Datasets}

In this section, we introduce the established \texttt{Reddit-v2} dataset and the public \texttt{Redial} dataset used for CRS model evaluations.

\begin{figure}[t]
\centering
\includegraphics[width=0.84\linewidth]{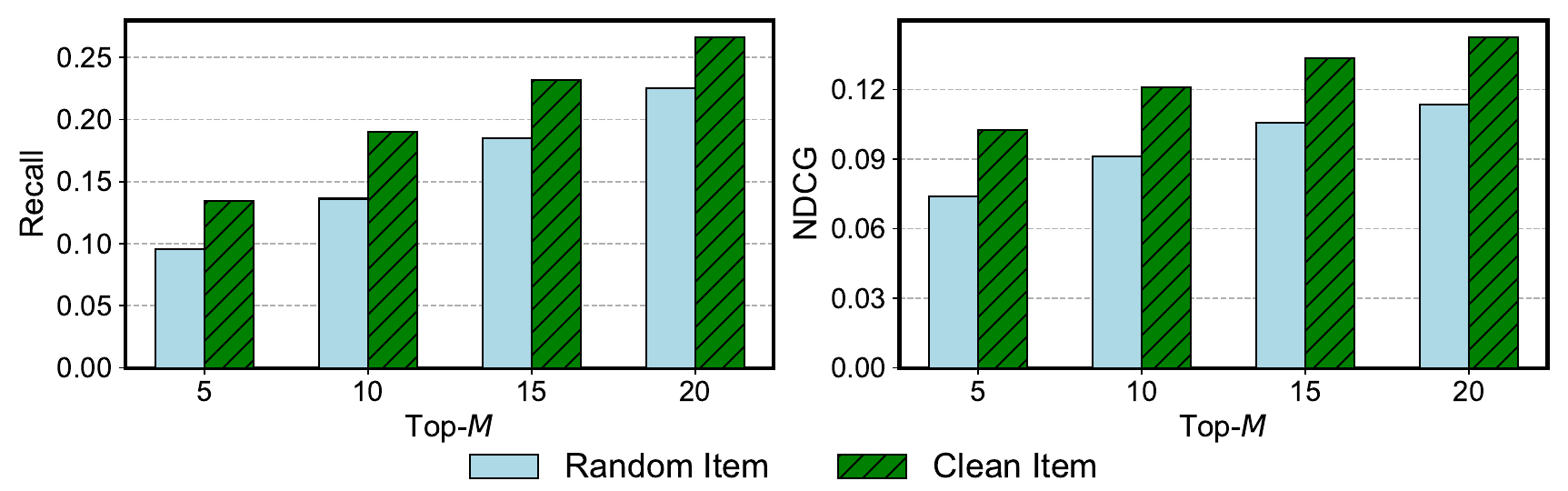}
\vspace{-3mm}
\caption{Comparison of zero-shot LLM on the Reddit-v2 dataset and the one with randomly replaced items.}
\vspace{-4mm}
\label{fig:noise}
\end{figure}

\subsubsection{\textbf{Reddit-v2 Dataset}}\label{sec:data} The largest real-world CRS dataset is the \texttt{Reddit} dataset established in \citep{he2023large}, which consists of dialogues collected from the Reddit website under movie-seeking topics. In each dialogue, the movie-seeker is treated as the \texttt{user}, whereas the responder is treated as the \texttt{system}. In addition, items (i.e., movies) were extracted from the utterances using a T5 model \cite{raffel2020exploring} fine-tuned on a simulated utterance-item dataset. However, due to the limited capacity of both the T5 model and the simulated training data, this strategy suffered from rather low accuracy in the extracted movies.

\begin{figure*}[t]
    \centering
    \begin{subfigure}{\textwidth}
        \centering
\includegraphics[width=0.71\textwidth]{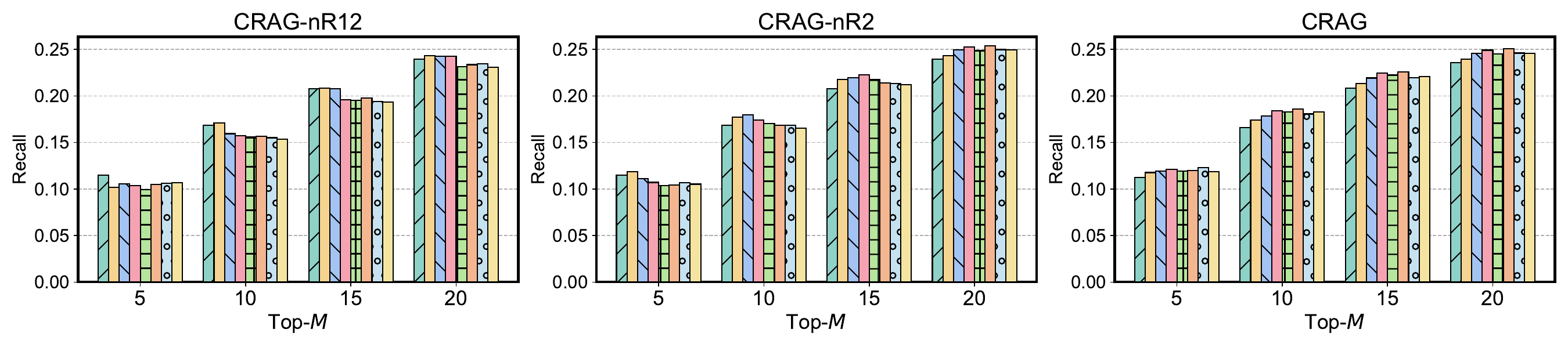} 
\vspace{-2mm}
        \caption{Reddit-v2 Dataset} 
    \end{subfigure}
    \hfill
    \begin{subfigure}{\textwidth}
        \centering        \includegraphics[width=0.71\textwidth]{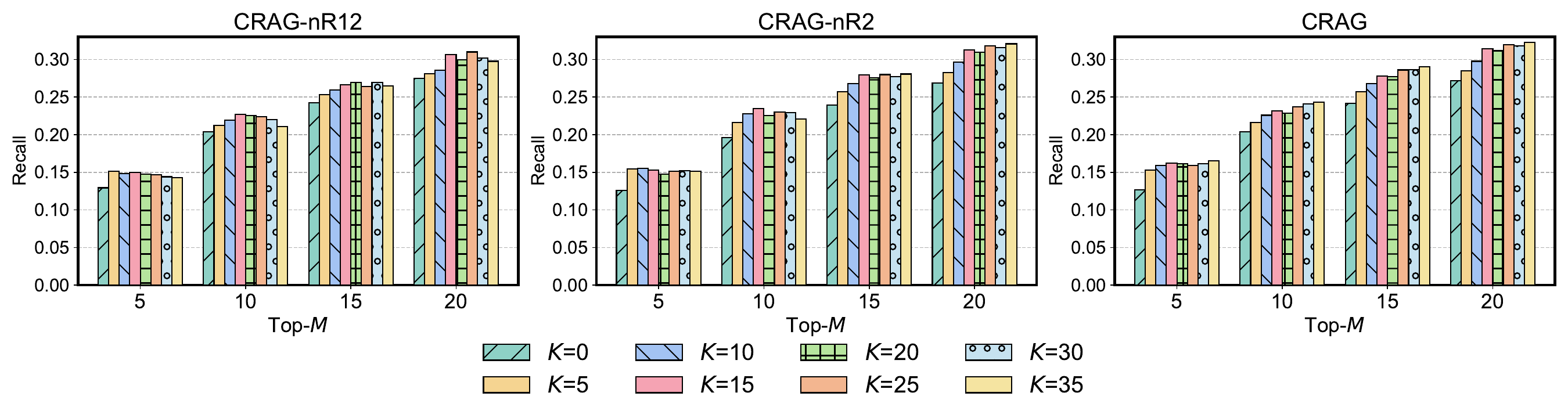}
        \vspace{-2mm}
        \caption{Redial Dataset}
        \vspace{-2mm}
    \end{subfigure}
    \vspace{-5mm}
    \caption{The influence of the number of items in the raw collaborative retrieval $K$ (depicted in different color) on the recommendation performance of CRAG-nR12, CRAG-nR2, and CRAG. X-axis denotes the recall evaluated at top-$M$ positions.}
    \label{fig:wrt_k}
    \vspace{-1mm}
\end{figure*}

To address the issue, we first refine the \texttt{Reddit} dataset (which we name \textbf{\textit{Reddit-v2}}) by extracting movies from the utterances with GPT-4o based on the strategy introduced in Section \ref{sec:entity}. To quantitatively verify the effectiveness of the movie name extraction, we reproduce the item-replacement experiment in \cite{he2023large}, where we compare the performance of a \texttt{Zero-shot LLM} before and after randomly replacing the extracted movies in the \texttt{Reddit-v2} dialogues. The results are illustrated in Fig. \ref{fig:noise}. In Fig. \ref{fig:noise} we can see a noticeable degradation in the recommendation performance when movies in the dialogues are randomly replaced ($\sim$0.05 for recall@5, where the difference of recall@5 is less than 0.01 in \cite{he2023large}). These results not only show that the refined \texttt{Reddit-v2} dataset is significantly cleaner, but also lead to \textbf{Finding 1}: \textit{items mentioned in the conversations play a critical role for LLMs to generate recommendations.}

\subsubsection{\textbf{Redial Dataset}} Another CRS dataset that we utilize in this paper is the \texttt{Redial} dataset, which is crowd-sourced from Amazon Mechanical Turk (AMT) by \citet{li2018towards} for movie recommendations. In the \texttt{Redial} dataset, movies mentioned in the utterances are tagged by the Turkers. Although this requirement eliminates the necessity of item recognition and database linking, it is not realistic in real-world applications. In addition, we found (as with \citet{he2023large}) that conversations in the \texttt{Redial} dataset can be overly polite, rigid, and succinct (e.g., replying with "Whatever Whatever I’m open to any suggestion." when being asked to clarify preferences), where the context information is comparatively insufficient. 

\subsection{Experimental Setup}

For the \texttt{Reddit-v2} dataset, we use the subset of dialogues in the last month (i.e., \textit{Dec. 2022}) as the test set, a subset from the prior month as the validation set, and all other dialogues as the training set (which are used for training traditional CRS baselines). Additionally, we establish the interaction data $\mathbf{R}$ based on the training dialogues, where each dialogue is treated as a pseudo-user $i$, and all the positively mentioned items are treated as the historical interactions $\mathbf{r}_{i}$. For both \texttt{Reddit-v2} and \texttt{Redial} datasets, we treat the set of mentioned items in all the dialogues as the item database $\mathcal{I}$ and all the items mentioned by the system as the catalog $\mathcal{Q}$. The statistics of both datasets are shown in Table \ref{tab:dataset} in the Appendix.

We consider two LLMs, i.e., GPT-4 and the latest GPT-4o, as the CRS backbone for \texttt{CRAG}\footnote{The experimental results with GPT-4 backbone are provided in Appendix \ref{sec:gpt4}.}. We excluded other LLM models, such as GPT-3.5 and GPT-3.5-turbo, due to their significantly weaker instruction-following abilities. Here, we note that for the evaluation of LLM-based CRSs, the choice of LLM faces an inherent trade-off between item coverage and data-leakage risk. For GPT-4, approximately 15\% of the movies in the item database $\mathcal{I}$ are released after its pretraining cut-off date, but all test dialogues are after its cut-off date. In contrast, GPT-4o covers all the items, but the test dialogues are before its cut-off date. However, even for GPT-4o, the risk of data leakage is low, as Reddit closed its crawling interface before GPT-4o. In addition, since the strongest baseline, i.e., the zero-shot LLM proposed in \cite{he2023large}, will use the same LLM as \texttt{CRAG}, the comparison remains fair despite the trade-off in LLM selections. 

\subsection{Analysis of the Two-step Reflections}\label{sec:exp}

In our experiments, we first analyze the key contribution of \texttt{CRAG}, i.e., the two-step reflection process defined in Eqs. (\ref{eq:ref_tre}), (\ref{eq:ref_rerank}), which improves the context-relevancy of collaborative retrieval and reranks the items in the recommendation list to prioritize more relevant ones. Specifically, we aim to explore when reflection works and how each reflection step contributes to the performance of \texttt{CRAG}.

\subsubsection{\textbf{Evaluation Setup}} To answer the above research questions, we design two variants of \texttt{CRAG}, i.e., \texttt{CRAG-nR12}, \texttt{CRAG-nR2}, and explore their performance when the number of items in the raw collaborative retrieval (i.e., $K$ in Eq. (\ref{eq:cr})) increases. Specifically, in \texttt{CRAG-nR12}, we removed both reflection steps, whereas in \texttt{CRAG-nR2}, we remove only the final reflect-and-rerank step. We note that when $K=0$, both \texttt{CRAG-nR12} and \texttt{CRAG-nR2} reduce to the zero-shot LLM as \cite{he2023large}. In addition, with the context-aware reflection module, the number of items actually getting augmented into the prompt could be less than $K$ for both \texttt{CRAG} and \texttt{CRAG-nR2}. In the recommendation step, all three models are asked to recommend \textit{\textbf{20}} movies.

\subsubsection{\textbf{Intra-variant Comparisons}} We first consider each \texttt{CRAG} variant \textit{separately} and focus on the performance trend when the number of raw retrievals $K$ increases. The results are shown in Fig. \ref{fig:wrt_k}. The bar group at $M$ shows the trend of recall@$M$ when $K$ increases from 0 to 35. In Fig. \ref{fig:wrt_k} we have three interesting findings:

\vspace{1mm}

\textbf{Finding 2.} \textit{Naive collaborative retrieval is not very effective} (see the \underline{leftmost} sub-figure). On the \texttt{Reddit-v2} dataset, the performance of \texttt{CRAG-nR12}, i.e., the \texttt{CRAG} variant without any reflection, generally \textit{decreases} when more items are retrieved and augmented into the prompt. This makes sense because the raw collaborative retrieval, i.e., $\mathcal{I}^{CR}_{k}$, does not consider any context in the dialogue, where context-irrelevant items bias the LLM’s recommendations. 

\vspace{1mm}

\textbf{Finding 3.} \textit{Context-aware reflection improves the coverage of relevant items but struggles with the item rank} (see the \underline{middle} sub-figure). This is reflected by the recall@20 bar group (as 20 is the number of movies we ask the LLM to recommend) for \texttt{CRAG-nR2}, where the metric now increases with a larger value of $K$ compared with \texttt{CRAG-nR12}. However, recall@5, 10 of \texttt{CRAG-nR2} quickly peak and then decrease as more items are retrieved. This suggests that, with $K$ growing, an increased number of relevant items are recommended by \texttt{CRAG-nR2}, but they may not be ranked in top positions.

\vspace{1mm}

\textbf{Finding 4.} \textit{Reflect-and-rerank addresses the rank bias and prioritizes most relevant items} (see the \underline{rightmost} sub-figure). For \texttt{CRAG} with both reflection steps, we note that recall@5, 10 also increase with growing $K$ (besides recall@20). This suggests that not only more relevant items are recommended by \texttt{CRAG}, but they are also increasingly getting ranked in the top (i.e., 5 and 10) positions. 

The above findings lead to the conclusion that \textit{LLMs are able to identify relevant items even if they cannot generate them.}  This is exemplified by the fact that for $K>0$, \texttt{CRAG} is able to exceed the performance of zero-shot LLMs (i.e., if $K=0$) and include an increased number of relevant items in the top-$M$ positions. As in \texttt{CRAG}, the LLM reflects on additional items generated by collaborative retrieval, this result suggests that the LLM is able to identify relevant items from among these additional items, even though the LLM was not able to generate these additional relevant items itself.

\begin{figure}[h]
    \centering
    \begin{subfigure}[b]{0.45\textwidth}
    \centering
   \includegraphics[width=0.99\linewidth]{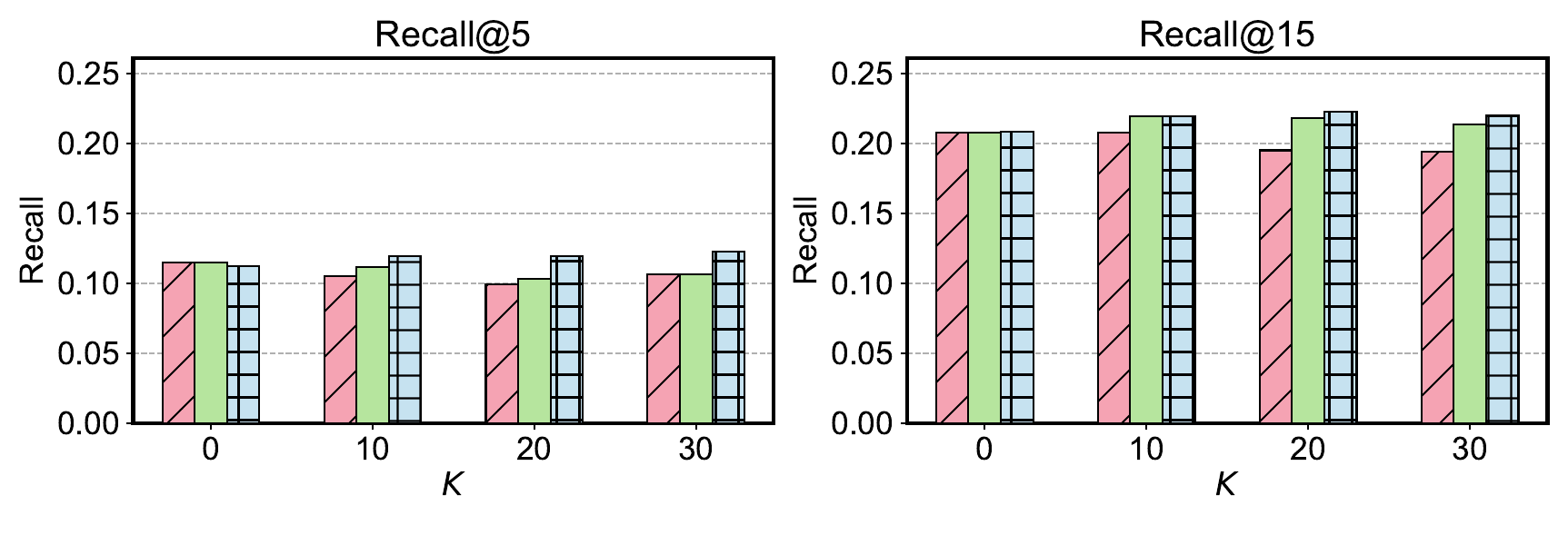}
    \vspace{-3mm}
    \caption{Reddit-v2 Dataset}
    \end{subfigure}
    
    \begin{subfigure}[b]{0.45\textwidth}
        \centering
\includegraphics[width=0.99\linewidth]{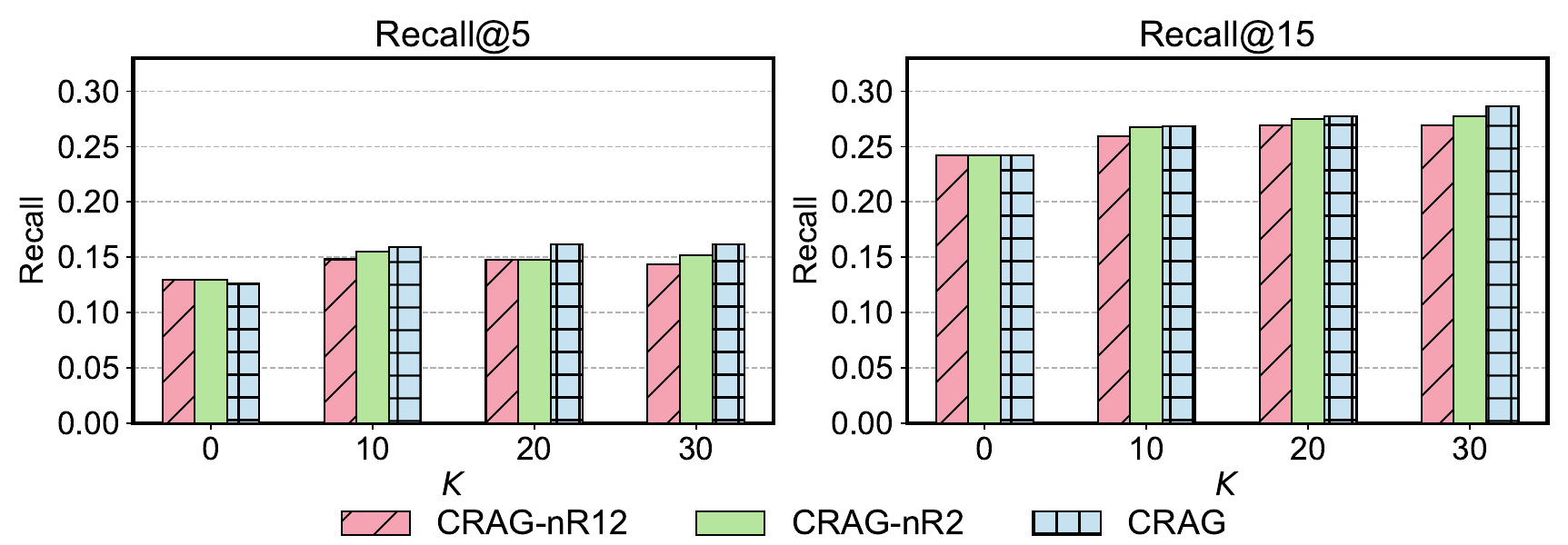}
        \caption{Redial Dataset}
    \end{subfigure}
    \vspace{-2mm}
    \caption{Comparison across different CRAG variants.}
    \vspace{-5mm}
    \label{fig:inter_wrt_k}
\end{figure}

\subsubsection{\textbf{Cross-variant Comparison}} In addition, we compare across the \texttt{CRAG} variants in Fig. \ref{fig:inter_wrt_k}, from which we can draw \textit{two more interesting conclusions} that are not evident in Fig. \ref{fig:wrt_k}: 

\textbf{Finding 5.} \textit{Self-reflection does not help.} We note that when $K=0$, \texttt{CRAG-nR12} and \texttt{CRAG-nR2} degenerate to the \texttt{zero-shot LLM}, and \texttt{CRAG} degenerates to adding self-reflect and rerank on the zero-shot generations. The left-most bar group in Fig. \ref{fig:inter_wrt_k} shows that when the recommendations are generated without external knowledge, self-reflection on the final recommendation list does not help. This makes sense, as for the \texttt{zero-shot LLM} model, the items reflected upon are generated based only on the same LLM's internal knowledge, where the self-reflection cannot introduce new knowledge.

\vspace{1mm}

\textbf{Finding 6.} \textit{Context is important for both reflection steps.} 
The larger improvement of \texttt{CRAG} over \texttt{CRAG-nR12} and \texttt{CRAG-nR2} on the \texttt{Reddit-v2} dataset compared with the \texttt{Redial} dataset shows that the two-step reflection works better for dialogues with richer context information (as for the \texttt{Reddit-v2} dataset). This shows that \texttt{CRAG} can effectively combine collaborative retrieval with the context-understanding ability of LLMs to improve LLM-based CRS.

\begin{figure}[t]
    \centering
    \begin{subfigure}{0.45\textwidth}
        \centering
        \includegraphics[width=\textwidth]{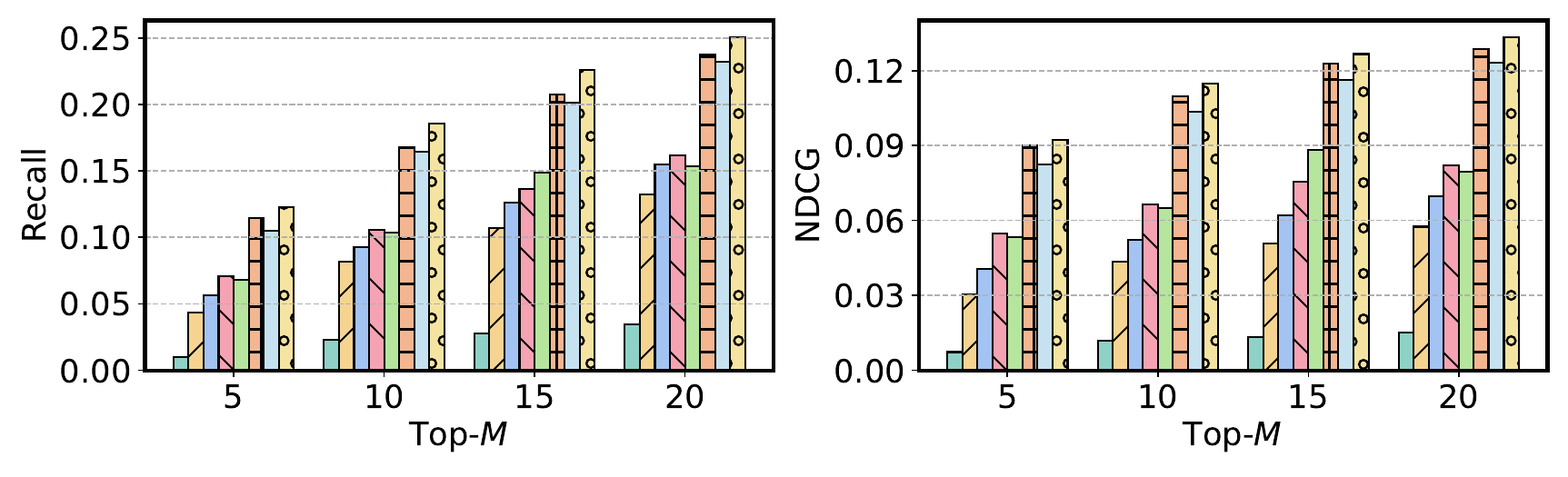} 
        \vspace{-5mm}    
        \caption{Reddit-v2 Dataset} 
    \end{subfigure}

    \vspace{2mm}

    \begin{subfigure}{0.45\textwidth}
        \centering
    \includegraphics[width=\textwidth]{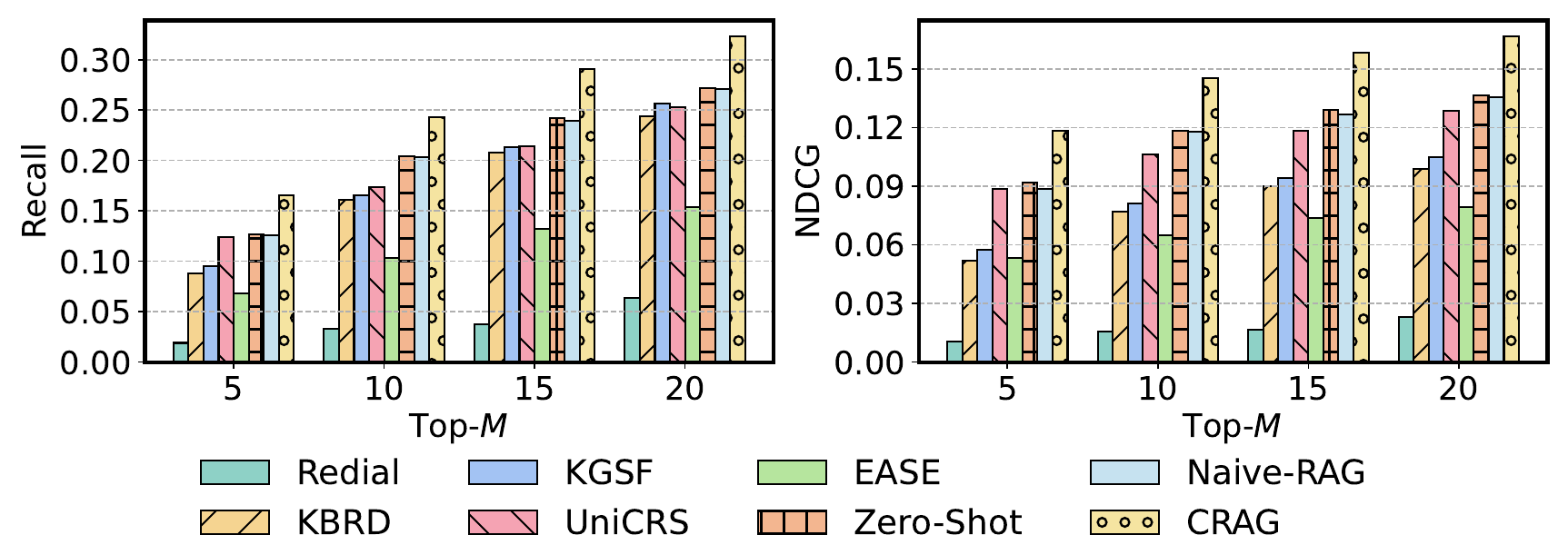}
        \caption{Redial Dataset}
    \end{subfigure}
\vspace{-3mm}    
    \caption{Comparison between CRAG and various baselines on the established Reddit-v2 and Redial datasets.}
    \label{fig:baseline}
    \vspace{-3mm}  
\end{figure}

\subsection{Comparison to Baselines}\label{sec:exp2}
In this section, we compare \texttt{CRAG} with various state-of-the-art RNN-, transformer-, and LLM-based CRS baselines as follows:
%
\begin{itemize}[leftmargin=0.5cm]
    \item \textbf{Redial} \cite{li2018towards} leverages a denoising autoencoder to model the mentioned items and to generate recommendations, while an RNN is used to model and generate conversations.
    \item \textbf{KBRD} \cite{chen2019towards} introduces a relational GNN (RGNN) on the DBpedia knowledge graph (KG) to model entities, and optimize similarity between co-occurring words and entities to fuse the semantics. 
    \item \textbf{KGSF} \cite{zhou2020improving} incorporates a word-level KG from ConceptNet to model the conversations and use mutual information maximization w.r.t. entity KG embeddings to fuse the entity information.
    \item \textbf{UniCRS} \cite{wang2022towards} introduces a pretrained transformer to capture the context information, with cross-attention \cite{vaswani2017attention} w.r.t. the entity KG embeddings (RGNN) used for semantic fusion. 
    \item \textbf{Zero-shot LLM} \cite{he2023large} directly inputs the dialogue with task-specific prompt and format instruction for CRS without any information retrieval from external knowledge databases.
    \item \textbf{Naive-RAG} \cite{lewis2020retrieval} retrieves item-related sentences from a database of movie plots and metadata based on the query-sentence semantic similarity to augment the zero-shot LLM.
\end{itemize}


For \texttt{Redial}, \texttt{KBRD}, and \texttt{KGSF}, we follow the implementation in \href{https://github.com/RUCAIBox/CRSLab/}{CRSLab}, where we adapt the evaluation codes (which replicate each conversation multiple times such that each has exactly one groundtruth) to make it consistent with the evaluation of \texttt{CRAG}. In addition, we limit the recommendations of all the baselines to the items in the catalog $\mathcal{Q}$. Finally, we include \textbf{EASE} \cite{steck2019embarrassingly} as a non-CRS baseline (as we adapt it for collaborative retrieval), whose recommendations are based only on items mentioned in the dialogue.

\begin{figure}[t]
    \centering
    \begin{subfigure}{0.45\textwidth}
        \centering
    \includegraphics[width=0.95\linewidth]{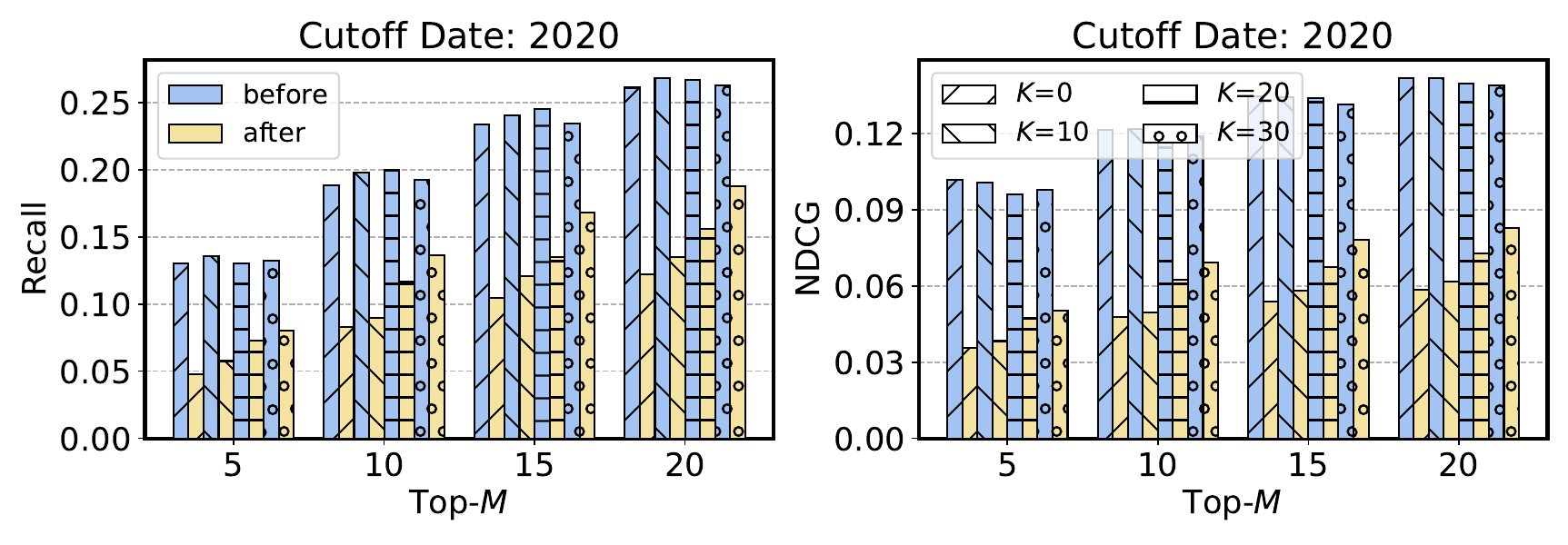}
        \vspace{-2mm}
        \caption{Reddit-v2 Dataset} 
    \end{subfigure}
    \begin{subfigure}{0.45\textwidth}
        \centering

          \vspace{2mm}
        \includegraphics[width=0.95\linewidth]{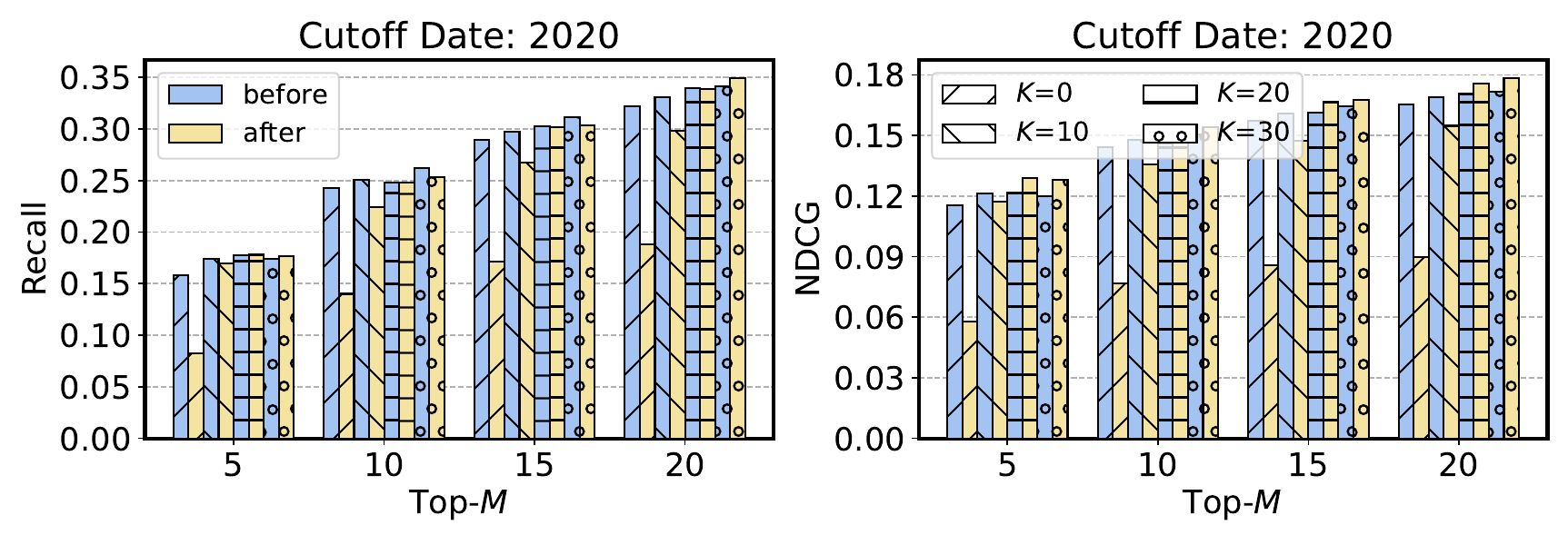}
        \vspace{-2mm}
        \caption{Redial Dataset}
    \end{subfigure}
    \vspace{-2mm}
    \caption{Comparison between the results on test dialogues separated by the release year of the groundtruth movies.}
    \vspace{-4mm}
    \label{fig:cutoff}
\end{figure}

The comparisons are shown in Fig. \ref{fig:baseline}, where we can see that the \texttt{Redial} model, which separately models and generates items and conversations, achieves the lowest performance. \texttt{KBRD} and \texttt{KGSF} improve over \texttt{redial} by introducing an external KG on the entities and strategies to fuse the entity and context semantics in the dialogue. \texttt{UniCRS} further leverages a pretrained transformer to model the context, which achieves the best performance among all the non-LLM-based baselines. However, due to the vast knowledge and reasoning ability of modern LLMs, Fig. \ref{fig:baseline} shows that a \texttt{Zero-shot LLM} improves substantially over the traditional methods. Regarding the RAG-based methods, we have the following findings:

\textbf{Finding 7.} Interestingly, we find that, \texttt{Naive-RAG}, which augments the \texttt{Zero-shot LLM} by retrieving relevant content/metadata as documents into the prompt, actually \textit{degrades} in performance. The reason could be the large semantic gap between words in the conversations and the implicit user preference. For example, for the dialogue in Fig. \ref{fig:framework}, most documents retrieved by \texttt{Naive-RAG} are from movies that directly have Brazil/Brazilian in the title, but the user mentioned Brazilian only as a quantifier to his/her true preferences, i.e., movies similar to \texttt{City of God} and \texttt{Bacurau}. 

\textbf{Finding 8.} \texttt{CRAG} achieves the best performance by all the metrics across both datasets compared with both \texttt{Zero-Shot LLM} and \texttt{Naive-RAG}, which further demonstrates the effectiveness of the collaborative retrieval with two-step reflection in \texttt{CRAG}.

\subsection{Evaluation w.r.t. the Recency of Items}\label{sec:abl}
In this section, we shed more light on the main effect that we identified for \texttt{CRAG}: while \texttt{CRAG} improves the recommendation accuracy for all cases, the gains are more substantial for movies that were released more recently. This is corroborated by the following experiment:
We first select a cut-off year (e.g., \texttt{2020}, but other years generally lead to similar results) and split the test dialogues into \texttt{before} and \texttt{after} groups: in the \texttt{before}-group, all groundtruth movies are released before the cut-off year, whereas in the \texttt{after}-group, at least one movie is released after the cut-off year. The results in Fig. \ref{fig:cutoff} show the following interesting findings:

\textbf{Finding 9.} \textit{LLMs are less effective in recommending more recent items.} This is reflected by the overall lower performance of \texttt{CRAG} on the \texttt{after}-group (yellow bars) than the \texttt{before}-group (blue bars).

\textbf{Finding 10.} \textit{\texttt{CRAG} leads to larger improvements for the recommendation of more recent items.} This is reflected by the larger metric increases for \texttt{CRAG} on the \texttt{after}-group with the growing of $K$, i.e., the number of items in the raw collaborative retrieval, compared to the \texttt{before}-group. Visually, this is reflected by the steeper metric improvements of the yellow bars (i.e., \texttt{after}-group) compared to the blue bars (i.e., \texttt{before}-group) when $K$ increases in Fig. \ref{fig:cutoff}.

\subsection{Retrieval and Recommendation}\label{sec:abl2}
\label{sec:pos}
Finally, we point out the importance of reflect-and-rerank step.
To this end, we examine the relation between items in the context-aware collaborative retrieval $\mathcal{I}^{aug}_{k}$ that get augmented into the prompt and the final list of items getting recommended by \texttt{CRAG-nR2} (i.e., $\mathcal{I}^{rec}_{k}$ in Eq. (\ref{eq:rec})) and \texttt{CRAG} (i.e., $\mathcal{I}^{r\&r}_{k}$ in Eq. (\ref{eq:ref_rerank})). Fig. \ref{fig:rank} shows the confusion matrix, where the element at row $i$ and column $j$ denotes the number of times when the $j$-th item in $\mathcal{I}^{aug}_{k}$ is put in the $i$-th position of recommendations (to save space, we selected $K=20$ and show only top 5 rows). The matrix at the top of Fig. \ref{fig:rank} shows the results of \texttt{CRAG-nR2}, which leads to the following findings:

 \begin{figure}[t]
\centering
\includegraphics[width=0.88\linewidth]{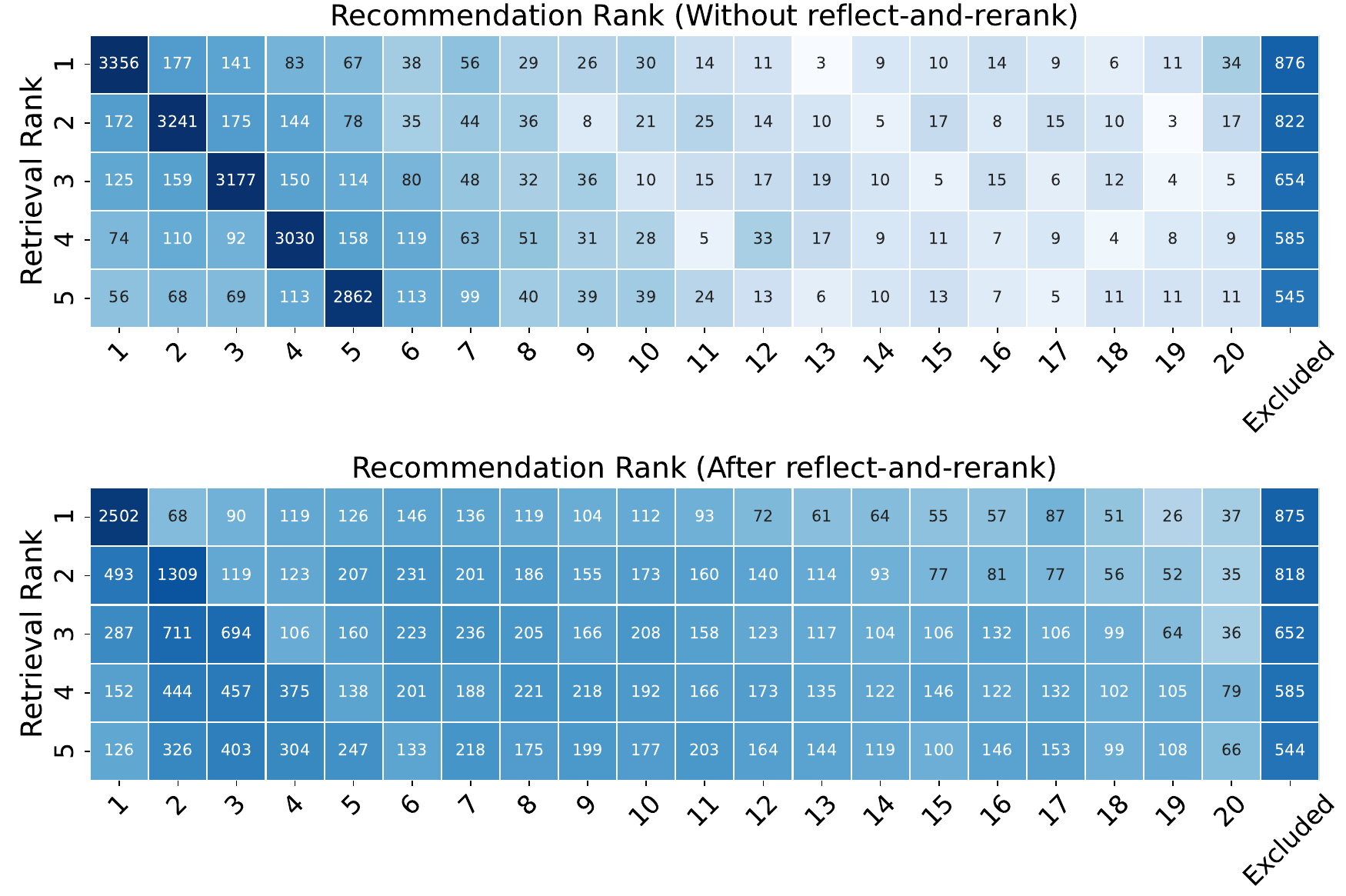}
\vspace{-5mm}
\caption{Confusion matrix for item rank in retrieval and recommendation w, w/o the reflect-and-rerank step.}
\vspace{-4mm}
\label{fig:rank}
\end{figure}

\textbf{Finding 11.} \textit{LLMs have the bias to replicate the retrieved items.} The dominating diagonal elements in the confusion matrix at the top of Fig. \ref{fig:rank} show that LLMs are indeed biased to replicating retrieved items at the beginning of the recommendation list. This is undesirable as the retrieved items only consider the CF information (as context-aware reflection only removes context-irrelevant items).

\textbf{Finding 12.} \textit{LLMs tend to replace items in the collaborative retrieval in place instead of removing them and filling in the next ones}. Otherwise, given the large number of items excluded from recommendations, the confusion matrix at the top of Fig. \ref{fig:rank} should have larger values for all lower-triangular elements (which denote the cases of upwardly lifted items from the retrieval $\mathcal{I}^{aug}_{k}$ to the recommendation list $\mathcal{I}^{rec}_{k}$) instead of only dominant diagonal elements.

With reflect-and-rerank introduced on the final recommendation list, the dominating diagonal elements vanish in the confusion-matrix for \texttt{CRAG} at the bottom of Fig. \ref{fig:rank}, which indicates that more relevant items are prioritized at the top of the recommendation list $\mathcal{I}^{r\&r}_{k}$, irrespective of whether these items were from collaborative retrieval $\mathcal{I}^{aug}_{k}$ or new ones generated by the LLM outside $\mathcal{I}^{aug}_{k}$.

\begin{figure}[t]
\centering
\includegraphics[width=0.88\linewidth]{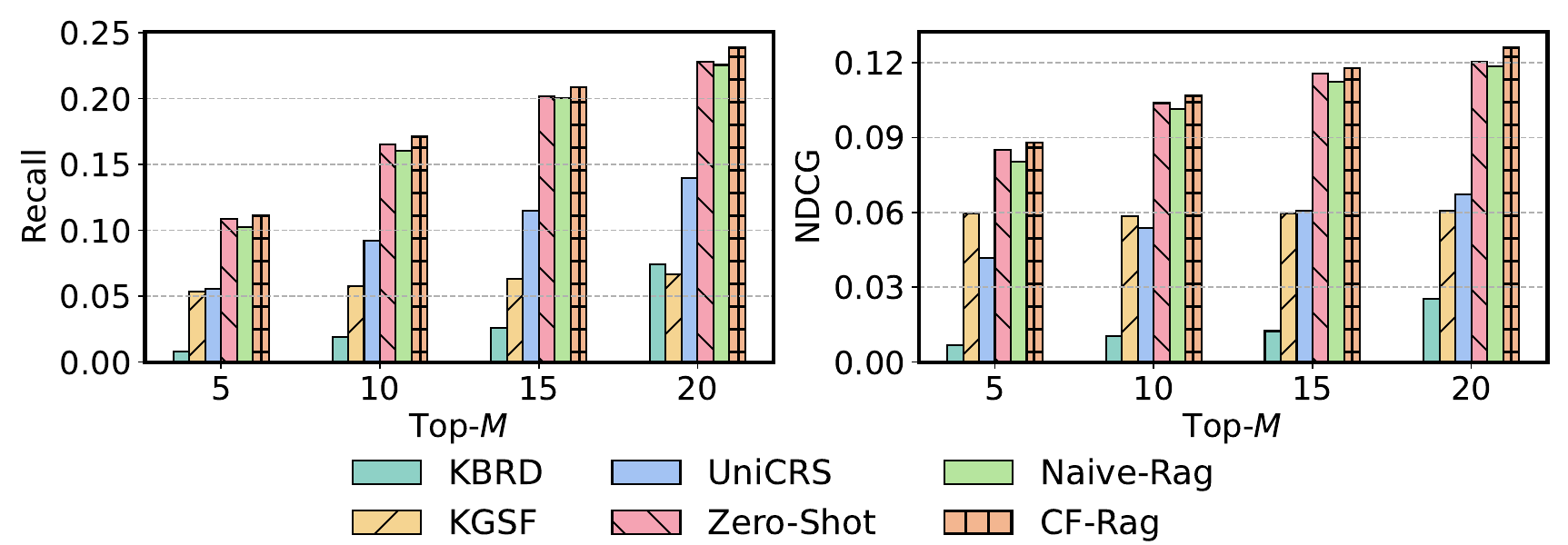}
\vspace{-3mm}
\caption{Comparison of CRAG to the baselines on the conversations with no item mentions in the Reddit-v2 dataset.}
\label{fig:baseline_cold}
\vspace{-3mm}
\end{figure}

\subsection{Conversations without Items Mentions}

In this section, we provide the results for \texttt{CRAG} on conversations with no explicitly mentioned items based on the pre-generation trick introduced in Section \ref{sec:coldstart}. We focus on the \texttt{Reddit-v2} dataset, as almost all the conversations in the \texttt{Redial} dataset contain at least one movie being mentioned. The comparison with baselines in Fig. \ref{fig:baseline_cold} leads to \textbf{Finding 13}: \textit{The relative performance of the baselines generally shows a similar trend as the cases where there are items mentioned in the conversation (see Fig. \ref{fig:baseline})}, although the improvement of \texttt{CRAG} over the \texttt{Zero-shot LLM} baseline is not as substantial.

\begin{figure}[t]
\centering
\includegraphics[width=0.9\linewidth]{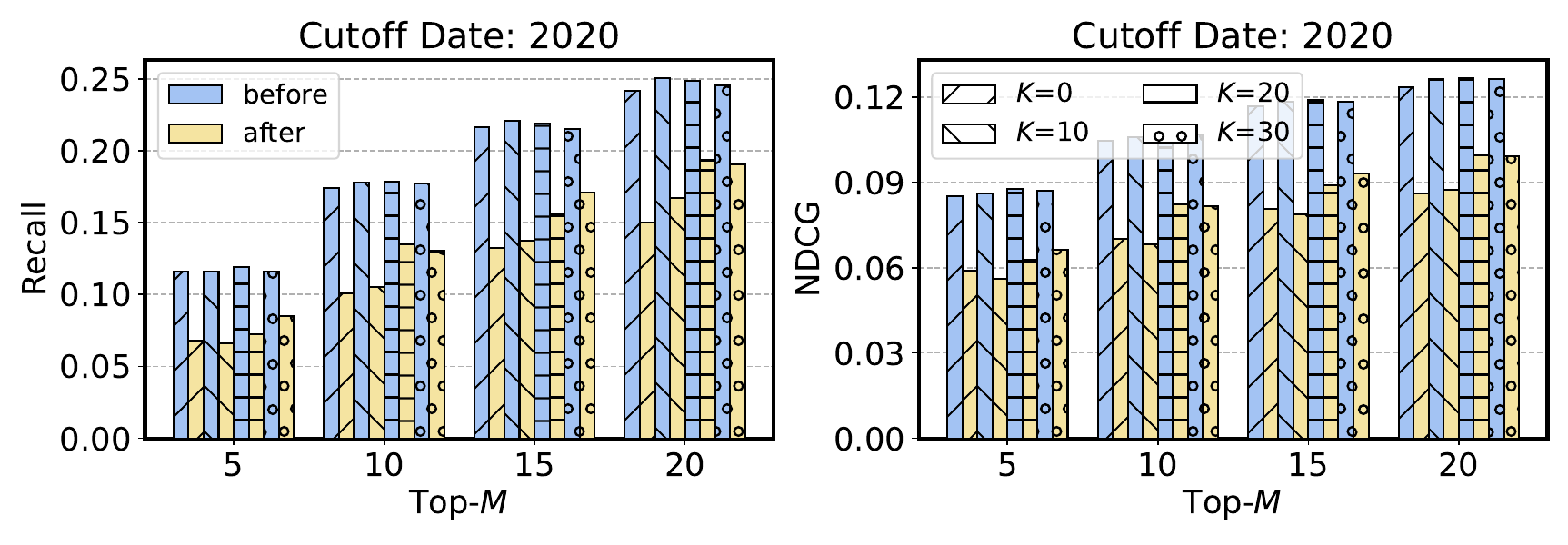}
\vspace{-4mm}
\caption{Results on Reddit-v2 where conversations are separated by the release year of the movies to be recommended.}
\vspace{-4mm}
\label{fig:year_cold}
\end{figure}

Analogous to the experiments in Fig. \ref{fig:cutoff}, the results for the conversations with no movie mentions are shown in 
 Fig. \ref{fig:year_cold}, which lead to \textbf{Finding 14}: Despite the smaller overall improvement of \texttt{CRAG} over \texttt{Zero-shot LLM} in the case where no items are mentioned in the dialogue, \textit{the improvement in the recommendation of movies with more recent release-years is still very evident}. This again demonstrates the increased effectiveness of \texttt{CRAG} over \texttt{zero-shot LLMs} in recommending movies that are more recently released.

\section{Conclusions}

In this paper, we proposed \texttt{CRAG}, the first approach that combines state-of-the-art, black-box LLMs with collaborative filtering for CRS. In our experiments, we showed that this results in improved recommendation accuracy on two publicly available movie conversational recommendation datasets, eclipsing the current state-of-the-art CRS methods, i.e., the zero-shot LLMs. We also provided several ablation studies to shed light on the inner workings of this approach. In particular, we found that the recently released movies benefited especially from \texttt{CRAG}. 
Apart from that, we established a refined version of the Reddit dataset on movie recommendations, where the extraction of movies mentioned in the dialogues is greatly improved. We also showed that this improvement in movie extraction accuracy can have a considerable impact on the derived insights. 


\begin{thebibliography}{44}


\ifx \showCODEN    \undefined \def \showCODEN     #1{\unskip}     \fi
\ifx \showDOI      \undefined \def \showDOI       #1{#1}\fi
\ifx \showISBNx    \undefined \def \showISBNx     #1{\unskip}     \fi
\ifx \showISBNxiii \undefined \def \showISBNxiii  #1{\unskip}     \fi
\ifx \showISSN     \undefined \def \showISSN      #1{\unskip}     \fi
\ifx \showLCCN     \undefined \def \showLCCN      #1{\unskip}     \fi
\ifx \shownote     \undefined \def \shownote      #1{#1}          \fi
\ifx \showarticletitle \undefined \def \showarticletitle #1{#1}   \fi
\ifx \showURL      \undefined \def \showURL       {\relax}        \fi
\providecommand\bibfield[2]{#2}
\providecommand\bibinfo[2]{#2}
\providecommand\natexlab[1]{#1}
\providecommand\showeprint[2][]{arXiv:#2}

\bibitem[Anthropic(2024)]%
        {anthropic2024}
\bibfield{author}{\bibinfo{person}{Anthropic}.} \bibinfo{year}{2024}\natexlab{}.
\newblock \bibinfo{title}{Claude 3.5 Sonnet}.
\newblock \bibinfo{howpublished}{\url{https://www.anthropic.com/news/claude-3-5-sonnet}}.
\newblock


\bibitem[Auer et~al\mbox{.}(2007)]%
        {auer2007dbpedia}
\bibfield{author}{\bibinfo{person}{S{\"o}ren Auer}, \bibinfo{person}{Christian Bizer}, \bibinfo{person}{Georgi Kobilarov}, \bibinfo{person}{Jens Lehmann}, \bibinfo{person}{Richard Cyganiak}, {and} \bibinfo{person}{Zachary Ives}.} \bibinfo{year}{2007}\natexlab{}.
\newblock \showarticletitle{Dbpedia: A nucleus for a web of open data}. In \bibinfo{booktitle}{\emph{International Semantic Web Conference}}. Springer, \bibinfo{pages}{722--735}.
\newblock


\bibitem[Bao et~al\mbox{.}(2023)]%
        {bao2023tallrec}
\bibfield{author}{\bibinfo{person}{Keqin Bao}, \bibinfo{person}{Jizhi Zhang}, \bibinfo{person}{Yang Zhang}, \bibinfo{person}{Wenjie Wang}, \bibinfo{person}{Fuli Feng}, {and} \bibinfo{person}{Xiangnan He}.} \bibinfo{year}{2023}\natexlab{}.
\newblock \showarticletitle{Tallrec: An effective and efficient tuning framework to align large language model with recommendation}. In \bibinfo{booktitle}{\emph{RecSys}}. \bibinfo{pages}{1007--1014}.
\newblock


\bibitem[Belghazi et~al\mbox{.}(2018)]%
        {belghazi2018mutual}
\bibfield{author}{\bibinfo{person}{Mohamed~Ishmael Belghazi}, \bibinfo{person}{Aristide Baratin}, \bibinfo{person}{Sai Rajeshwar}, \bibinfo{person}{Sherjil Ozair}, \bibinfo{person}{Yoshua Bengio}, \bibinfo{person}{Aaron Courville}, {and} \bibinfo{person}{Devon Hjelm}.} \bibinfo{year}{2018}\natexlab{}.
\newblock \showarticletitle{Mutual information neural estimation}. In \bibinfo{booktitle}{\emph{ICML}}. \bibinfo{pages}{531--540}.
\newblock


\bibitem[Chen et~al\mbox{.}(2019)]%
        {chen2019towards}
\bibfield{author}{\bibinfo{person}{Qibin Chen}, \bibinfo{person}{Junyang Lin}, \bibinfo{person}{Yichang Zhang}, \bibinfo{person}{Ming Ding}, \bibinfo{person}{Yukuo Cen}, \bibinfo{person}{Hongxia Yang}, {and} \bibinfo{person}{Jie Tang}.} \bibinfo{year}{2019}\natexlab{}.
\newblock \showarticletitle{Towards Knowledge-Based Recommender Dialog System}. In \bibinfo{booktitle}{\emph{EMNLP}}.
\newblock


\bibitem[Chung et~al\mbox{.}(2014)]%
        {chung2014empirical}
\bibfield{author}{\bibinfo{person}{Junyoung Chung}, \bibinfo{person}{Caglar Gulcehre}, \bibinfo{person}{Kyunghyun Cho}, {and} \bibinfo{person}{Yoshua Bengio}.} \bibinfo{year}{2014}\natexlab{}.
\newblock \showarticletitle{Empirical evaluation of gated recurrent neural networks on sequence modeling}. In \bibinfo{booktitle}{\emph{NeurIPS Workshop on Deep Learning}}.
\newblock


\bibitem[Daiber et~al\mbox{.}(2013)]%
        {daiber2013improving}
\bibfield{author}{\bibinfo{person}{Joachim Daiber}, \bibinfo{person}{Max Jakob}, \bibinfo{person}{Chris Hokamp}, {and} \bibinfo{person}{Pablo~N Mendes}.} \bibinfo{year}{2013}\natexlab{}.
\newblock \showarticletitle{Improving efficiency and accuracy in multilingual entity extraction}. In \bibinfo{booktitle}{\emph{Semantic}}. \bibinfo{pages}{121--124}.
\newblock


\bibitem[Feng et~al\mbox{.}(2023)]%
        {feng2023large}
\bibfield{author}{\bibinfo{person}{Yue Feng}, \bibinfo{person}{Shuchang Liu}, \bibinfo{person}{Zhenghai Xue}, \bibinfo{person}{Qingpeng Cai}, \bibinfo{person}{Lantao Hu}, \bibinfo{person}{Peng Jiang}, \bibinfo{person}{Kun Gai}, {and} \bibinfo{person}{Fei Sun}.} \bibinfo{year}{2023}\natexlab{}.
\newblock \showarticletitle{A large language model enhanced conversational recommender system}.
\newblock \bibinfo{journal}{\emph{arXiv preprint arXiv:2308.06212}} (\bibinfo{year}{2023}).
\newblock


\bibitem[Gao et~al\mbox{.}(2021)]%
        {gao2021advances}
\bibfield{author}{\bibinfo{person}{Chongming Gao}, \bibinfo{person}{Wenqiang Lei}, \bibinfo{person}{Xiangnan He}, \bibinfo{person}{Maarten de Rijke}, {and} \bibinfo{person}{Tat-Seng Chua}.} \bibinfo{year}{2021}\natexlab{}.
\newblock \showarticletitle{Advances and challenges in conversational recommender systems: A survey}.
\newblock \bibinfo{journal}{\emph{AI Open}}  \bibinfo{volume}{2} (\bibinfo{year}{2021}), \bibinfo{pages}{100--126}.
\newblock


\bibitem[Gao et~al\mbox{.}(2023)]%
        {gao2023retrieval}
\bibfield{author}{\bibinfo{person}{Yunfan Gao}, \bibinfo{person}{Yun Xiong}, \bibinfo{person}{Xinyu Gao}, \bibinfo{person}{Kangxiang Jia}, \bibinfo{person}{Jinliu Pan}, \bibinfo{person}{Yuxi Bi}, \bibinfo{person}{Yi Dai}, \bibinfo{person}{Jiawei Sun}, {and} \bibinfo{person}{Haofen Wang}.} \bibinfo{year}{2023}\natexlab{}.
\newblock \showarticletitle{Retrieval-augmented generation for large language models: A survey}.
\newblock \bibinfo{journal}{\emph{arXiv preprint arXiv:2312.10997}} (\bibinfo{year}{2023}).
\newblock


\bibitem[Gu et~al\mbox{.}(2016)]%
        {gu2016incorporating}
\bibfield{author}{\bibinfo{person}{J Gu}, \bibinfo{person}{Z Lu}, \bibinfo{person}{H Li}, {and} \bibinfo{person}{VOK Li}.} \bibinfo{year}{2016}\natexlab{}.
\newblock \showarticletitle{Incorporating copying mechanism in sequence-to-sequence learning}. In \bibinfo{booktitle}{\emph{ACL}}.
\newblock


\bibitem[Ham et~al\mbox{.}(2020)]%
        {ham2020end}
\bibfield{author}{\bibinfo{person}{Donghoon Ham}, \bibinfo{person}{Jeong-Gwan Lee}, \bibinfo{person}{Youngsoo Jang}, {and} \bibinfo{person}{Kee-Eung Kim}.} \bibinfo{year}{2020}\natexlab{}.
\newblock \showarticletitle{End-to-end neural pipeline for goal-oriented dialogue systems using GPT-2}. In \bibinfo{booktitle}{\emph{ACL}}. \bibinfo{pages}{583--592}.
\newblock


\bibitem[He et~al\mbox{.}(2023)]%
        {he2023large}
\bibfield{author}{\bibinfo{person}{Zhankui He}, \bibinfo{person}{Zhouhang Xie}, \bibinfo{person}{Rahul Jha}, \bibinfo{person}{Harald Steck}, \bibinfo{person}{Dawen Liang}, \bibinfo{person}{Yesu Feng}, \bibinfo{person}{Bodhisattwa~Prasad Majumder}, \bibinfo{person}{Nathan Kallus}, {and} \bibinfo{person}{Julian McAuley}.} \bibinfo{year}{2023}\natexlab{}.
\newblock \showarticletitle{Large language models as zero-shot conversational recommenders}. In \bibinfo{booktitle}{\emph{CIKM}}.
\newblock


\bibitem[Hua et~al\mbox{.}(2023)]%
        {hua2023index}
\bibfield{author}{\bibinfo{person}{Wenyue Hua}, \bibinfo{person}{Shuyuan Xu}, \bibinfo{person}{Yingqiang Ge}, {and} \bibinfo{person}{Yongfeng Zhang}.} \bibinfo{year}{2023}\natexlab{}.
\newblock \showarticletitle{How to index item ids for recommendation foundation models}. In \bibinfo{booktitle}{\emph{SIGIR}}. \bibinfo{pages}{195--204}.
\newblock


\bibitem[Jannach et~al\mbox{.}(2021)]%
        {jannach2021survey}
\bibfield{author}{\bibinfo{person}{Dietmar Jannach}, \bibinfo{person}{Ahtsham Manzoor}, \bibinfo{person}{Wanling Cai}, {and} \bibinfo{person}{Li Chen}.} \bibinfo{year}{2021}\natexlab{}.
\newblock \showarticletitle{A survey on conversational recommender systems}.
\newblock \bibinfo{journal}{\emph{ACM Computing Surveys (CSUR)}} \bibinfo{volume}{54}, \bibinfo{number}{5} (\bibinfo{year}{2021}), \bibinfo{pages}{1--36}.
\newblock


\bibitem[Jannach et~al\mbox{.}(2010)]%
        {jannach2010recommender}
\bibfield{author}{\bibinfo{person}{Dietmar Jannach}, \bibinfo{person}{Markus Zanker}, \bibinfo{person}{Alexander Felfernig}, {and} \bibinfo{person}{Gerhard Friedrich}.} \bibinfo{year}{2010}\natexlab{}.
\newblock \bibinfo{booktitle}{\emph{Recommender Systems: An Introduction}}.
\newblock \bibinfo{publisher}{Cambridge University Press}.
\newblock


\bibitem[Kim et~al\mbox{.}(2024)]%
        {kim2024large}
\bibfield{author}{\bibinfo{person}{Sein Kim}, \bibinfo{person}{Hongseok Kang}, \bibinfo{person}{Seungyoon Choi}, \bibinfo{person}{Donghyun Kim}, \bibinfo{person}{Minchul Yang}, {and} \bibinfo{person}{Chanyoung Park}.} \bibinfo{year}{2024}\natexlab{}.
\newblock \showarticletitle{Large language models meet collaborative filtering: an efficient all-round LLM-based recommender system}. In \bibinfo{booktitle}{\emph{KDD}}. \bibinfo{pages}{1395--1406}.
\newblock


\bibitem[Koren et~al\mbox{.}(2021)]%
        {koren2021advances}
\bibfield{author}{\bibinfo{person}{Yehuda Koren}, \bibinfo{person}{Steffen Rendle}, {and} \bibinfo{person}{Robert Bell}.} \bibinfo{year}{2021}\natexlab{}.
\newblock \showarticletitle{Advances in collaborative filtering}.
\newblock \bibinfo{journal}{\emph{Recommender Systems Handbook}} (\bibinfo{year}{2021}), \bibinfo{pages}{91--142}.
\newblock

\bibitem[Lewis et~al\mbox{.}(2020)]%
        {lewis2020retrieval}
\bibfield{author}{\bibinfo{person}{Patrick Lewis}, \bibinfo{person}{Ethan Perez}, \bibinfo{person}{Aleksandra Piktus}, \bibinfo{person}{Fabio Petroni}, \bibinfo{person}{Vladimir Karpukhin}, \bibinfo{person}{Naman Goyal}, \bibinfo{person}{Heinrich K{\"u}ttler}, \bibinfo{person}{Mike Lewis}, \bibinfo{person}{Wen-tau Yih}, \bibinfo{person}{Tim Rockt{\"a}schel}, {et~al\mbox{.}}} \bibinfo{year}{2020}\natexlab{}.
\newblock \showarticletitle{Retrieval-augmented generation for knowledge-intensive NLP tasks}. In \bibinfo{booktitle}{\emph{NeurIPS}}. \bibinfo{pages}{9459--9474}.
\newblock


\bibitem[Li et~al\mbox{.}(2018)]%
        {li2018towards}
\bibfield{author}{\bibinfo{person}{Raymond Li}, \bibinfo{person}{Samira Ebrahimi~Kahou}, \bibinfo{person}{Hannes Schulz}, \bibinfo{person}{Vincent Michalski}, \bibinfo{person}{Laurent Charlin}, {and} \bibinfo{person}{Chris Pal}.} \bibinfo{year}{2018}\natexlab{}.
\newblock \showarticletitle{Towards deep conversational recommendations}. In \bibinfo{booktitle}{\emph{NeurIPS}}.
\newblock


\bibitem[Liang et~al\mbox{.}(2018)]%
        {liang2018variational}
\bibfield{author}{\bibinfo{person}{Dawen Liang}, \bibinfo{person}{Rahul~G Krishnan}, \bibinfo{person}{Matthew~D Hoffman}, {and} \bibinfo{person}{Tony Jebara}.} \bibinfo{year}{2018}\natexlab{}.
\newblock \showarticletitle{Variational autoencoders for collaborative filtering}. In \bibinfo{booktitle}{\emph{WWW}}. \bibinfo{pages}{689--698}.
\newblock


\bibitem[Mnih and Salakhutdinov(2007)]%
        {mnih2007probabilistic}
\bibfield{author}{\bibinfo{person}{Andriy Mnih} {and} \bibinfo{person}{Russ~R Salakhutdinov}.} \bibinfo{year}{2007}\natexlab{}.
\newblock \showarticletitle{Probabilistic matrix factorization}. In \bibinfo{booktitle}{\emph{NeurIPS}}, Vol.~\bibinfo{volume}{20}.
\newblock

\vfill\eject

\bibitem[OpenAI(2024)]%
        {openai2024}
\bibfield{author}{\bibinfo{person}{OpenAI}.} \bibinfo{year}{2024}\natexlab{}.
\newblock \bibinfo{title}{Hello GPT-4o}.
\newblock \bibinfo{howpublished}{\url{https://openai.com/index/hello-gpt-4o/}}.
\newblock


\bibitem[Radford et~al\mbox{.}(2019)]%
        {radford2019language}
\bibfield{author}{\bibinfo{person}{Alec Radford}, \bibinfo{person}{Jeffrey Wu}, \bibinfo{person}{Rewon Child}, \bibinfo{person}{David Luan}, \bibinfo{person}{Dario Amodei}, \bibinfo{person}{Ilya Sutskever}, {et~al\mbox{.}}} \bibinfo{year}{2019}\natexlab{}.
\newblock \showarticletitle{Language models are unsupervised multitask learners}.
\newblock \bibinfo{journal}{\emph{OpenAI blog}} \bibinfo{volume}{1}, \bibinfo{number}{8} (\bibinfo{year}{2019}), \bibinfo{pages}{9}.
\newblock


\bibitem[Raffel et~al\mbox{.}(2020)]%
        {raffel2020exploring}
\bibfield{author}{\bibinfo{person}{Colin Raffel}, \bibinfo{person}{Noam Shazeer}, \bibinfo{person}{Adam Roberts}, \bibinfo{person}{Katherine Lee}, \bibinfo{person}{Sharan Narang}, \bibinfo{person}{Michael Matena}, \bibinfo{person}{Yanqi Zhou}, \bibinfo{person}{Wei Li}, {and} \bibinfo{person}{Peter~J Liu}.} \bibinfo{year}{2020}\natexlab{}.
\newblock \showarticletitle{Exploring the limits of transfer learning with a unified text-to-text transformer}.
\newblock \bibinfo{journal}{\emph{JMLR}} \bibinfo{volume}{21}, \bibinfo{number}{140} (\bibinfo{year}{2020}), \bibinfo{pages}{1--67}.
\newblock


\bibitem[Ren et~al\mbox{.}(2024)]%
        {ren2024representation}
\bibfield{author}{\bibinfo{person}{Xubin Ren}, \bibinfo{person}{Wei Wei}, \bibinfo{person}{Lianghao Xia}, \bibinfo{person}{Lixin Su}, \bibinfo{person}{Suqi Cheng}, \bibinfo{person}{Junfeng Wang}, \bibinfo{person}{Dawei Yin}, {and} \bibinfo{person}{Chao Huang}.} \bibinfo{year}{2024}\natexlab{}.
\newblock \showarticletitle{Representation learning with large language models for recommendation}. In \bibinfo{booktitle}{\emph{WWW}}. \bibinfo{pages}{3464--3475}.
\newblock


\bibitem[Rendle(2010)]%
        {rendle2010factorization}
\bibfield{author}{\bibinfo{person}{Steffen Rendle}.} \bibinfo{year}{2010}\natexlab{}.
\newblock \showarticletitle{Factorization machines}. In \bibinfo{booktitle}{\emph{ICDM}}. \bibinfo{pages}{995--1000}.
\newblock


\bibitem[Speer et~al\mbox{.}(2017)]%
        {speer2017conceptnet}
\bibfield{author}{\bibinfo{person}{Robyn Speer}, \bibinfo{person}{Joshua Chin}, {and} \bibinfo{person}{Catherine Havasi}.} \bibinfo{year}{2017}\natexlab{}.
\newblock \showarticletitle{Conceptnet 5.5: An open multilingual graph of general knowledge}. In \bibinfo{booktitle}{\emph{AAAI}}, Vol.~\bibinfo{volume}{31}.
\newblock


\bibitem[Steck(2019a)]%
        {steck2019high}
\bibfield{author}{\bibinfo{person}{Harald Steck}.} \bibinfo{year}{2019}\natexlab{a}.
\newblock \showarticletitle{Collaborative Filtering via High-Dimensional Regression}.
\newblock \bibinfo{journal}{\emph{arXiv preprint arXiv:1904.13033}} (\bibinfo{year}{2019}).
\newblock


\bibitem[Steck(2019b)]%
        {steck2019embarrassingly}
\bibfield{author}{\bibinfo{person}{Harald Steck}.} \bibinfo{year}{2019}\natexlab{b}.
\newblock \showarticletitle{Embarrassingly shallow autoencoders for sparse data}. In \bibinfo{booktitle}{\emph{WWW}}. \bibinfo{pages}{3251--3257}.
\newblock


\bibitem[Sun and Zhang(2018)]%
        {sun2018conversational}
\bibfield{author}{\bibinfo{person}{Yueming Sun} {and} \bibinfo{person}{Yi Zhang}.} \bibinfo{year}{2018}\natexlab{}.
\newblock \showarticletitle{Conversational recommender system}. In \bibinfo{booktitle}{\emph{SIGIR}}. \bibinfo{pages}{235--244}.
\newblock


\bibitem[Vaswani et~al\mbox{.}(2017)]%
        {vaswani2017attention}
\bibfield{author}{\bibinfo{person}{Ashish Vaswani}, \bibinfo{person}{Noam Shazeer}, \bibinfo{person}{Niki Parmar}, \bibinfo{person}{Jakob Uszkoreit}, \bibinfo{person}{Llion Jones}, {and} \bibinfo{person}{Aidan~N Gomez}.} \bibinfo{year}{2017}\natexlab{}.
\newblock \showarticletitle{Attention is all you need}. In \bibinfo{booktitle}{\emph{NeurIPS}}.
\newblock


\bibitem[Vincent et~al\mbox{.}(2008)]%
        {vincent2008extracting}
\bibfield{author}{\bibinfo{person}{Pascal Vincent}, \bibinfo{person}{Hugo Larochelle}, \bibinfo{person}{Yoshua Bengio}, {and} \bibinfo{person}{Pierre-Antoine Manzagol}.} \bibinfo{year}{2008}\natexlab{}.
\newblock \showarticletitle{Extracting and composing robust features with denoising autoencoders}. In \bibinfo{booktitle}{\emph{ICML}}. \bibinfo{pages}{1096--1103}.
\newblock


\bibitem[Wang et~al\mbox{.}(2022)]%
        {wang2022towards}
\bibfield{author}{\bibinfo{person}{Xiaolei Wang}, \bibinfo{person}{Kun Zhou}, \bibinfo{person}{Ji-Rong Wen}, {and} \bibinfo{person}{Wayne~Xin Zhao}.} \bibinfo{year}{2022}\natexlab{}.
\newblock \showarticletitle{Towards unified conversational recommender systems via knowledge-enhanced prompt learning}. In \bibinfo{booktitle}{\emph{KDD}}. \bibinfo{pages}{1929--1937}.

\bibitem[Wei et~al\mbox{.}(2024)]%
        {wei2024llmrec}
\bibfield{author}{\bibinfo{person}{Wei Wei}, \bibinfo{person}{Xubin Ren}, \bibinfo{person}{Jiabin Tang}, \bibinfo{person}{Qinyong Wang}, \bibinfo{person}{Lixin Su}, \bibinfo{person}{Suqi Cheng}, \bibinfo{person}{Junfeng Wang}, \bibinfo{person}{Dawei Yin}, {and} \bibinfo{person}{Chao Huang}.} \bibinfo{year}{2024}\natexlab{}.
\newblock \showarticletitle{LLMRec: Large language models with graph augmentation for recommendation}. In \bibinfo{booktitle}{\emph{WSDM}}. \bibinfo{pages}{806--815}.
\newblock


\bibitem[Wu et~al\mbox{.}(2024a)]%
        {wu2024coral}
\bibfield{author}{\bibinfo{person}{Junda Wu}, \bibinfo{person}{Cheng-Chun Chang}, \bibinfo{person}{Tong Yu}, \bibinfo{person}{Zhankui He}, \bibinfo{person}{Jianing Wang}, \bibinfo{person}{Yupeng Hou}, {and} \bibinfo{person}{Julian McAuley}.} \bibinfo{year}{2024}\natexlab{a}.
\newblock \showarticletitle{CoRAL: Collaborative Retrieval-Augmented Large Language Models Improve Long-tail Recommendation}. In \bibinfo{booktitle}{\emph{KDD}}. \bibinfo{pages}{3391--3401}.
\newblock


\bibitem[Wu et~al\mbox{.}(2024b)]%
        {wu2024survey}
\bibfield{author}{\bibinfo{person}{Likang Wu}, \bibinfo{person}{Zhi Zheng}, \bibinfo{person}{Zhaopeng Qiu}, \bibinfo{person}{Hao Wang}, \bibinfo{person}{Hongchao Gu}, \bibinfo{person}{Tingjia Shen}, \bibinfo{person}{Chuan Qin}, \bibinfo{person}{Chen Zhu}, \bibinfo{person}{Hengshu Zhu}, \bibinfo{person}{Qi Liu}, {et~al\mbox{.}}} \bibinfo{year}{2024}\natexlab{b}.
\newblock \showarticletitle{A survey on large language models for recommendation}.
\newblock \bibinfo{journal}{\emph{World Wide Web}} \bibinfo{volume}{27}, \bibinfo{number}{5} (\bibinfo{year}{2024}), \bibinfo{pages}{60}.
\newblock

\bibitem[Xi et~al\mbox{.}(2024a)]%
        {xi2024towards}
\bibfield{author}{\bibinfo{person}{Yunjia Xi}, \bibinfo{person}{Weiwen Liu}, \bibinfo{person}{Jianghao Lin}, \bibinfo{person}{Xiaoling Cai}, \bibinfo{person}{Hong Zhu}, \bibinfo{person}{Jieming Zhu}, \bibinfo{person}{Bo Chen}, \bibinfo{person}{Ruiming Tang}, \bibinfo{person}{Weinan Zhang}, {and} \bibinfo{person}{Yong Yu}.} \bibinfo{year}{2024}\natexlab{a}.
\newblock \showarticletitle{Towards open-world recommendation with knowledge augmentation from large language models}. In \bibinfo{booktitle}{\emph{RecSys}}. \bibinfo{pages}{12--22}.
\newblock


\bibitem[Xi et~al\mbox{.}(2024b)]%
        {xi2023towards}
\bibfield{author}{\bibinfo{person}{Yunjia Xi}, \bibinfo{person}{Weiwen Liu}, \bibinfo{person}{Jianghao Lin}, \bibinfo{person}{Xiaoling Cai}, \bibinfo{person}{Hong Zhu}, \bibinfo{person}{Jieming Zhu}, \bibinfo{person}{Bo Chen}, \bibinfo{person}{Ruiming Tang}, \bibinfo{person}{Weinan Zhang}, \bibinfo{person}{Rui Zhang}, {et~al\mbox{.}}} \bibinfo{year}{2024}\natexlab{b}.
\newblock \showarticletitle{Towards open-world recommendation with knowledge augmentation from large language models}. In \bibinfo{booktitle}{\emph{RecSys}}.
\newblock


\bibitem[Zhao et~al\mbox{.}(2024)]%
        {zhao2023survey}
\bibfield{author}{\bibinfo{person}{Wayne~Xin Zhao}, \bibinfo{person}{Kun Zhou}, \bibinfo{person}{Junyi Li}, \bibinfo{person}{Tianyi Tang}, \bibinfo{person}{Xiaolei Wang}, \bibinfo{person}{Yupeng Hou}, \bibinfo{person}{Yingqian Min}, \bibinfo{person}{Beichen Zhang}, \bibinfo{person}{Junjie Zhang}, \bibinfo{person}{Zican Dong}, {et~al\mbox{.}}} \bibinfo{year}{2024}\natexlab{}.
\newblock \showarticletitle{A survey of large language models}.
\newblock \bibinfo{journal}{\emph{arXiv preprint arXiv:2303.18223}} (\bibinfo{year}{2024}).
\newblock


\bibitem[Zheng et~al\mbox{.}(2024)]%
        {zheng2024adapting}
\bibfield{author}{\bibinfo{person}{Bowen Zheng}, \bibinfo{person}{Yupeng Hou}, \bibinfo{person}{Hongyu Lu}, \bibinfo{person}{Yu Chen}, \bibinfo{person}{Wayne~Xin Zhao}, \bibinfo{person}{Ming Chen}, {and} \bibinfo{person}{Ji-Rong Wen}.} \bibinfo{year}{2024}\natexlab{}.
\newblock \showarticletitle{Adapting large language models by integrating collaborative semantics for recommendation}. In \bibinfo{booktitle}{\emph{ICDE}}. \bibinfo{pages}{1435--1448}.
\newblock


\bibitem[Zhou et~al\mbox{.}(2020)]%
        {zhou2020improving}
\bibfield{author}{\bibinfo{person}{Kun Zhou}, \bibinfo{person}{Wayne~Xin Zhao}, \bibinfo{person}{Shuqing Bian}, \bibinfo{person}{Yuanhang Zhou}, \bibinfo{person}{Ji-Rong Wen}, {and} \bibinfo{person}{Jingsong Yu}.} \bibinfo{year}{2020}\natexlab{}.
\newblock \showarticletitle{Improving conversational recommender systems via knowledge graph based semantic fusion}. In \bibinfo{booktitle}{\emph{KDD}}. \bibinfo{pages}{1006--1014}.
\newblock


\bibitem[Zhou et~al\mbox{.}(2022)]%
        {zhou2022c2}
\bibfield{author}{\bibinfo{person}{Yuanhang Zhou}, \bibinfo{person}{Kun Zhou}, \bibinfo{person}{Wayne~Xin Zhao}, \bibinfo{person}{Cheng Wang}, \bibinfo{person}{Peng Jiang}, {and} \bibinfo{person}{He Hu}.} \bibinfo{year}{2022}\natexlab{}.
\newblock \showarticletitle{C$^2$-CRS: Coarse-to-fine contrastive learning for conversational recommender system}. In \bibinfo{booktitle}{\emph{WSDM}}. \bibinfo{pages}{1488--1496}.
\newblock


\bibitem[Zhu et~al\mbox{.}(2024)]%
        {zhu2024collaborative}
\bibfield{author}{\bibinfo{person}{Yaochen Zhu}, \bibinfo{person}{Liang Wu}, \bibinfo{person}{Qi Guo}, \bibinfo{person}{Liangjie Hong}, {and} \bibinfo{person}{Jundong Li}.} \bibinfo{year}{2024}\natexlab{}.
\newblock \showarticletitle{Collaborative large language model for recommender systems}. In \bibinfo{booktitle}{\emph{WWW}}. \bibinfo{pages}{3162--3172}.
\newblock


\end{thebibliography}


\newpage

\appendix

\definecolor{movie}{rgb}{1,1,0}

\noindent {\huge \textbf{Appendix}}

\vspace{3mm}

\noindent In the appendix, we discuss the related work of \texttt{CRAG} (see Section \ref{sec:rel_work}), provide detailed analysis and statistics of the established \texttt{Reddit-v2} dataset (see Section \ref{sec:data}), provide the details of the prompts used in the main paper (see Section \ref{sec:prompts}), and provide additional experimental results of \texttt{CRAG} with GPT-4 backbone (see Section \ref{sec:gpt4}).

\section{Related Work}
\label{sec:rel_work}

In this section, we review the related work of \texttt{CRAG}, which includes both conversational recommender systems and research on large language models (LLM) with collaborative filtering.

\subsection{Conversational Recommender Systems}

Conversational recommender systems (CRS) aim to generate recommendations through natural language interactions with users \cite{jannach2021survey,gao2021advances}. Throughout the dialogue with users, there usually exist two types of information, i.e., \textit{item} and \textit{context}, where the latter denotes the non-item words that users utter alongside the items to express their preferences. To handle these two aspects, CRS models generally contain the following two phases: \textbf{\textit{(i)}} \textit{modeling}, which learns to understand both items and context mentioned in the dialogue, and \textbf{\textit{(ii)}} \textit{generation}, which generates items and words in natural language based on the dialogue understanding as the response. 

From the \textit{modeling} perspective, a typical CRS involves three key components: entity modeling, context modeling, and semantic fusion. Various traditional recommendation models, such as factorization machines \citep{rendle2010factorization} and denoising autoencoders \citep{vincent2008extracting}, have been used to model the items mentioned in the dialogues \citep{li2018towards, chen2019towards}. Context modeling, on the other hand, utilizes language models with recurrent neural networks (RNNs) \citep{chung2014empirical} or transformers \citep{vaswani2017attention, ham2020end} to capture the conversational flow and background information. To integrate item and context information, semantic fusion techniques such as mutual information maximization \citep{belghazi2018mutual} and cross-attention mechanisms \citep{vaswani2017attention} have been employed \citep{zhou2020improving, wang2022towards} to comprehensively understand the user preferences. In addition, knowledge databases, such as DBPedia \cite{auer2007dbpedia} and ConceptNet \cite{speer2017conceptnet}, have been used to enhance both item and context modeling with external information. 

For the \textit{generation} phase, early methods introduced a switch mechanism, i.e., a binary predictor, to decide whether the next token to be generated should be a word or an item \citep{li2018towards, chen2019towards}. Afterward, approaches such as copy mechanism \citep{gu2016incorporating} are used to align item and word tokens in the same generation space. Recently, \citet{wang2022towards} introduced an \texttt{<item>} token when generating the context, which enables the system to comprehensively consider the generated context for item recommendations. \textit{The advent of large language models (LLMs) has further blurred the boundaries between items and context, as well as between modeling and generation phases of CRSs}. LLMs possess extensive knowledge and reasoning abilities, allowing them to understand items and context simultaneously in the form of natural language. Moreover, the generation of items and context in the responses can be unified in the textual space, leveraging the LLM's capacity to produce coherent natural language outputs.

However, LLMs are comparatively less effective in the recommendation of more recent items due to fewer relevant documents in the training corpora. In addition, LLMs struggle to leverage collaborative filtering knowledge, which is highly informative for recommendations. These two challenges motivate us to introduce collaborative retrieval with two-step reflection in \texttt{CRAG} to augment the LLM's recommendations with context-aware CF knowledge. 

\subsection{LLM with Collaborative Filtering}

Recently, recommender system researchers have recognized the importance of integrating collaborative filtering (CF) with large language models (LLMs) to further enhance their recommendation abilities \citep{wu2024survey}. Most works focus on the \textit{white-box} LLMs, where the model weights are accessible to the researchers. One promising strategy is to introduce new tokens for users/items to capture the collaborative filtering knowledge. These tokens can be independently assigned for each user/item \citep{zhu2024collaborative,bao2023tallrec} or clustered based on semantic indexing \citep{hua2023index}. The embeddings associated with the user/item tokens can be learned with language modeling on natural language sequences converted from user-item interactions \citep{zhu2024collaborative} or predicted from pretrained CF models based on external neural networks \citep{kim2024large, zheng2024adapting}. While white-box LLMs provide the possibility to introduce CF knowledge through model finetuning, they are generally smaller in scale compared to large proprietary LLMs. Due to the inaccessibility of model weights, combining CF with \textit{black-box} LLMs is less explored. One strategy is to augment CF models with LLMs' analysis of user preferences \citep{ren2024representation,wei2024llmrec,xi2024towards}. To use the LLM itself as the recommender, \citet{wu2024coral} proposed to transform user-item interactions into the prompt for LLMs to understand the user preference and utilize a policy network to reduce the redundancy. However, these approaches focus on traditional symmetric CF settings, which are not suitable for CRS with asymmetric item mention/recommendation and complex contextual information.

\texttt{CRAG} distinguishes itself by effectively combining CF with black-box, state-of-the-art LLMs that comprehensively consider the interaction data and the context in the dialogue for recommendations. By introducing context-aware retrieval and a two-step reflection process, \texttt{CRAG} addresses the limitations of zero-shot LLM-based CRS and substantially enhances the recommendation quality.

\begin{figure}[t]
\centering
\includegraphics[width=0.42\textwidth]{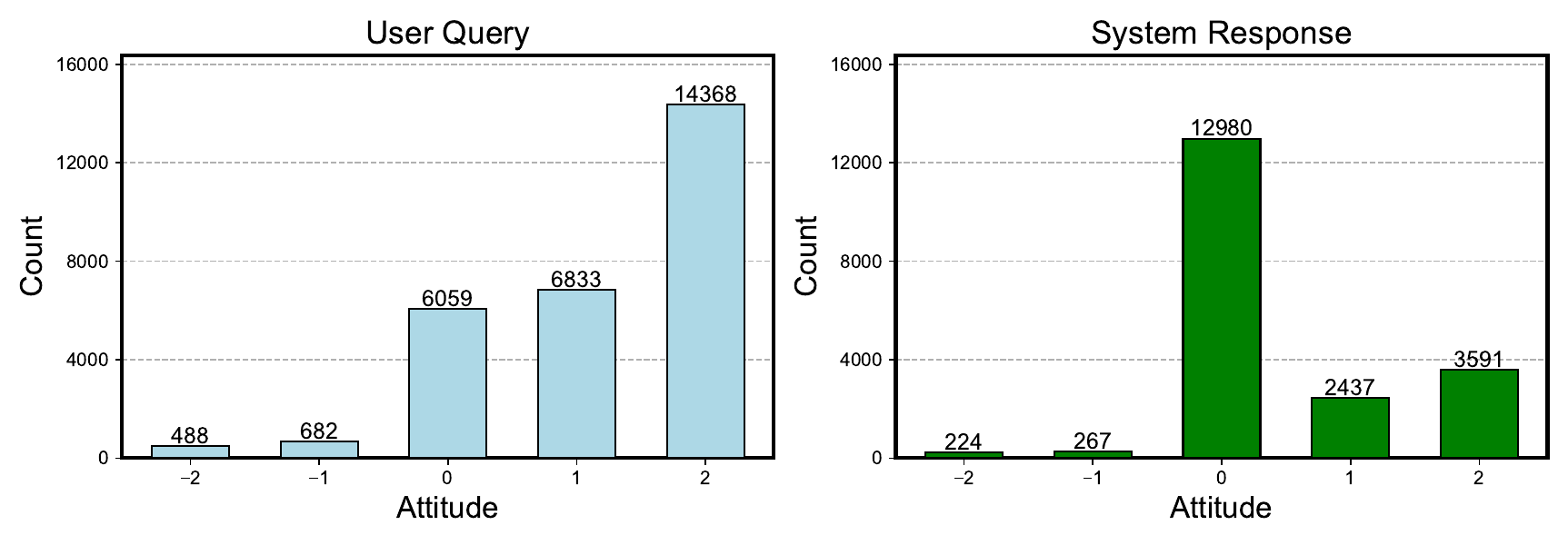} 
\vspace{-3mm}    
\caption{Distribution of attitudes for movie mentions in user queries and system responses for the Reddit-v2 test set.} 
\vspace{-5mm}   
\label{fig:attitude}
\end{figure}

\section{Details of the Reddit-v2 Dataset}

In this section, we provide details of the established \texttt{Reddit-v2} dataset. Specifically, we provide qualitative analysis of the movie name and attitude extraction, and various related dataset statistics.

\subsection{Comparison with Original Reddit Dataset}
\label{sec:ent_noise}

We first present the comparison results of movie name extraction between \texttt{Reddit-v2} and the original Reddit dataset in Tables \ref{tab:noisy vs clean context}, \ref{tab:noisy vs clean response}. As shown above, with the proposed LLM-based entity link and bi-level match and reflection strategy, \texttt{Reddit-v2} is more accurate in extracting movie names. Based on Tables \ref{tab:noisy vs clean context}, \ref{tab:noisy vs clean response}, we provide a detailed analysis of the reasons why the original Reddit dataset fails to accurately extract movie names from user queries and system responses, which can be summarized into three cases as follows:

\vspace{1mm}
\noindent \textbf{\textit{(i)}} \textbf{\textit{First, we note that user queries can be noisy, with movies being misspelled or abbreviated.}} Without a comprehensive understanding of the context, it becomes difficult for the trained T5-based entity recognition model to accurately identify the correct movie names. For example, in the 501-\textit{st} example, the user query states:

\vspace{2mm}
\begin{mdframed}[backgroundcolor=black!10] 
"...I feel like since the COVID lockdown I’ve seen every sci-fi action movie of this millennium... Things in the vein of the more modern \colorbox{movie!30}{AvP} movies, Battle of LA, the Frank Grillo and his son fighting aliens series that I’m blanking on the name of, Pacific Rim franchise, etc." 
\end{mdframed}
\vspace{2mm}
In this instance, the user uses "AvP" to refer to \texttt{Alien vs. Predator}, yet the original Reddit dataset fails to extract the correct title. 

\vspace{1mm}
\noindent \textbf{\textit{(ii)}} \textbf{\textit{In addition, we note that certain movie titles are ambiguous}} and can blend into the context of the user query, making it challenging for the model to distinguish them from the context. 
For example, in the 1092-\textit{nd} example, the user query states as follows:
\vspace{2mm}
\begin{mdframed}[backgroundcolor=black!10] 
"Can you suggest some Netflix series for people who are really alone... For instance, I was watching the new \colorbox{movie!30}{Wednesday} series and hoping I could relate to \colorbox{movie!30}{Wednesday} Addams..."
\end{mdframed}
\vspace{2mm}
Without prior knowledge of the series \texttt{Wednesday}, the entity recognition model might mistakenly interpret "Wednesday" in the user query as a reference to a day of the week rather than the title of a show. This makes it challenging to correctly identify \texttt{Wednesday} in the context. Similarly, in the 155-\textit{th} example, the query reads, 
\vspace{2mm}
\begin{mdframed}[backgroundcolor=black!10]
"...I have been looking for movies based on small American towns... The only movie that comes to mind is \colorbox{movie!30}{It}..."
\end{mdframed}
\vspace{2mm}
Here, even without recognizing that \texttt{It} refers to a specific movie, the sentence remains semantically coherent. In both instances, accurate name extraction relies heavily on GPT-4o's knowledge of the relevant movies and the ability to understand nuanced context.

\vspace{2mm}

\noindent \textbf{\textit{(iii)}} \textbf{\textit{Finally, we note sometimes the exact movie names mentioned by the user can be ambiguous}}. In such cases, identifying the optimal movie names relies heavily on the reasoning capabilities of the entity recognition model, which is typically achievable only by large language models like GPT-4o. For example, in the 105-\textit{th} example in the \texttt{Reddit-v2} test dataset, user query states:
\vspace{2mm}
\begin{mdframed}[backgroundcolor=black!10]
"...It gets mentioned a lot here, but Amelie is a movie that always lifts me up. This year I’d also recommend \colorbox{movie!30}{Everything,} \colorbox{movie!30}{Everywhere, All at Once}".
\end{mdframed}
\vspace{2mm}
The original Reddit dataset mistakenly recognizes the highlighted part "Everything,
Everywhere, All at Once" in the query into three movies—\texttt{Everything}, \texttt{Everywhere}, and \texttt{All at Once}. However, based on the context, it can be inferred that the user means the Oscar-winning film \texttt{Everything Everywhere All at Once}. 

\subsection{Analysis of Attitude Extraction}

We then qualitatively analyze the attitude extracted alongside the movie names in the \texttt{Reddit-v2} dataset. The results are provided in Table \ref{tab: positive}, Table \ref{tab: neutral}, and Table \ref{tab: negative}, which correspond to the examples of positive, neutral, and negative attitudes, respectively.

When users mention a movie either in their queries or as recommendations, they often convey a personal attitude toward it. In most cases, the LLM effectively infers whether the user holds a positive or negative sentiment toward the movies based on the surrounding context. In the 519-\textit{th} example, the user query states:
\vspace{2mm}
\begin{mdframed}[backgroundcolor=black!10] 
"Best Foreign Movies? I recently watched \colorbox{movie!30}{Troll} and \colorbox{movie!30}{Pan's} \colorbox{movie!30}{Labyrinth}. I wasn’t always fond of movies with subtitles, but I really enjoy them now. What are some good Sci-fi/Fantasy foreign films?" 
\end{mdframed}
\vspace{2mm}
In this case, the LLM rates the user's attitude toward the two mentioned movies as a 2, indicating a very positive attitude. Another straightforward example is the 879-\textit{th}, where the user writes, 
\vspace{2mm}
\begin{mdframed}[backgroundcolor=black!10] 
"...Movies like The \colorbox{movie!30}{Hangover, Superbad} are just so stale and overrated. Any suggestions, please? I need a good laugh tonight."
\end{mdframed}
\vspace{2mm}
Here, the \texttt{Hangover} and \texttt{Superbad} are rated as -2, reflecting the user's clearly negative attitude toward them. We also observe that if movies are recommended in earlier stages of the conversation but the user does not express any clear attitude toward them, they are usually assigned a rating of 0, indicating a neutral stance. For instance, in the 1418-\textit{th} example, the conversation is as follows: 
\vspace{2mm}
\begin{mdframed}[backgroundcolor=black!10] 
USER: [Request] Feel good movies?; SYSTEM: \colorbox{movie!30}{Rescued by} \colorbox{movie!30}{Ruby}; USER: Gonna give this one a go right now, thanks!" 
\end{mdframed}
\vspace{2mm}
In this case, the LLM rates the user's attitude toward \texttt{Rescued by Ruby} as 0, reflecting the user's neutral attitude.

In cases where the user's attitude is mixed, the LLM can discern subtle nuances and read between the lines. For example, in the 1353-\textit{rd} example, the user writes the following in the query: 
\vspace{2mm}
\begin{mdframed}[backgroundcolor=black!10] 
"Movies with interracial relationships, that aren’t strictly ABOUT that? So not stuff like \colorbox{movie!30}{Jungle Fever, Get Out}, etc." 
\end{mdframed}
\vspace{2mm}
Here, the users' attitudes toward \texttt{Jungle Fever} and \texttt{Get Out} are judged as -1, as the user does not express a strongly negative attitude but indicates that these movies do not align with their request. 

\section{Statistics of the Reddit-v2 Dataset}
\label{sec:data}

To address item extraction noise in the original Reddit dataset  \citep{he2023large}, we run the LLM-based entity extraction strategy introduced in Section \ref{sec:entity} to establish the refined \texttt{Reddit-v2} dataset. For evaluation, we select the subset of dialogues with start date on the last month (i.e., \textit{Dec. 2022}) as the test set, where the meta information (as with the \texttt{Redial} dataset) is illustrated in Table \ref{tab:dataset}. In addition, the distribution of attitudes for user query and system response is illustrated in Fig. \ref{fig:attitude}. The large number of attitudes $0$ for system response is due to succinct recommendations with only movie names, where the attitude is difficult to judge by LLM. Therefore, $0$ is also treated as a positive attitude for the system responses.

\begin{table}[t]
\centering
\caption{Statistics of the Reddit-v2 and Redial \underline{test} sets, where \textbf{\#Conv.} denotes the number of test samples with items mentioned in the dialogue, and \#Conv. (X) denotes the number of test samples with NO items mentioned in the dialogue.}
\vspace{-3mm}
\begin{tabular}{l|cccc}
\hline
\hline
\textbf{Dataset}      & \textbf{\#Conv.} & \textbf{\#Conv. (X)}  & \textbf{\#Items} & \textbf{\#Catalog} \\
\hline
\hline
\texttt{Reddit-v2}                    & 5,613     & 2,231         & 5,384        & 4,752   \\
\texttt{Redial} \cite{li2018towards}  & 2,998       & 619          & 1,915        & 1,476   \\
\hline
\hline
\end{tabular}
\label{tab:dataset}
\end{table}

\begin{figure*}[t]
    \centering
        \centering
\includegraphics[width=0.75\textwidth]{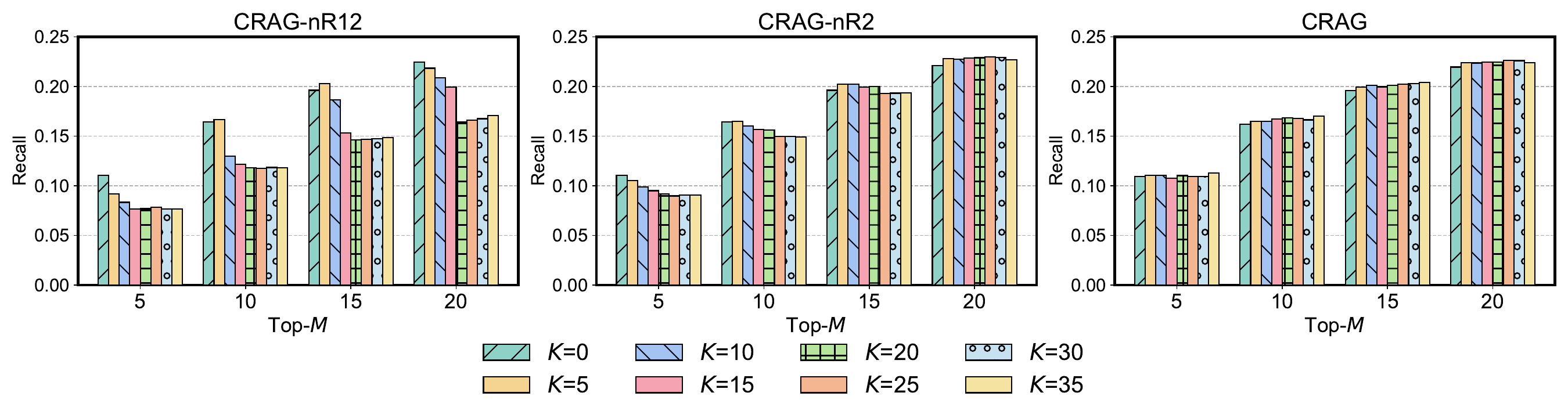} 
\vspace{-2mm}
        \caption{The influence of the number of items in the raw collaborative retrieval $K$ (depicted by bars with different colors) on the recommendation performance of CRAG-nR12, CRAG-nR2, and CRAG (all with GPT-4 backbone). X-axis denotes the recall evaluated at top-$M$ positions.} 
    \label{fig:wrt_k_4}
\end{figure*}

\section{Prompts Used in the Main Paper}
\label{sec:prompts}

In this section, we provide the task-specific prompts and format instructions that we defined in the main paper for the prompting and reflection process of \texttt{CRAG} (see Fig. \ref{fig:framework}) as follows:

\vspace{2mm}

\noindent \textbf{Eq. (\ref{eq:ent_ext_raw}): LLM-based Entity Extraction}

\vspace{2mm}
\begin{mdframed}[backgroundcolor=black!10] 

{\color{red}$T_{e}$}: Pretend you are a movie recommender system. You (a recommender system) will be given a user's query that seeks movie recommendations. Based on the query, you need to extract movie names mentioned in the user's query and analyze the user's attitude toward each movie. You need to reply with standardized movie names (with grammatical errors corrected and abbreviations fixed), as well as the user's attitude toward the movie. 

\vspace{3mm}
\noindent {\color{red}$F_{e}$}: Specifically, the movie names need to be formatted in the IMDB style, with the year bracketed if possible (do not add the year if you are not sure). In addition, the attitude is represented in one of [-2, -1, 0, 1, 2], where -2 stands for very negative, -1 stands for negative, 0 stands for neutral, 1 stands for positive, and 2 stands for very positive. You need to reply with the number as an attitude instead of the textual description. If there are movie names mentioned in the query, list each movie name and the user's attitude (number in -2 to 2) in the form of movie\_name\#\#\#\#attitude, where different movies are listed in different lines with no extra sentences. 
Reply NO if no movie names are mentioned in the query. 

\vspace{2mm}
\noindent  {\color{red}$s_{t}$}: Here is the user's query: \{\}.
\end{mdframed}

\vspace{2mm}

\noindent\hrulefill

\vspace{2mm}
\noindent \textbf{Eq. (\ref{eq:ent_ext_ref}): Reflection on Bi-level Matched Entities}

\vspace{3mm}
\begin{mdframed}[backgroundcolor=black!10] 
\noindent {\color{red}$T^{ref}_{e}$}: Pretend you are a movie recommender system. You, as the recommender system, will be given part of the dialogue between a user seeking a movie recommendation and yourself, along with the extracted movie names (which may potentially be incorrect). Even if the extracted movie names are correct, the wording might not be precise. Therefore, you will be provided with the best match for each extracted movie name from an external database using (1) character-level fuzzy match and (2) word-level BM25 match (a space will be provided if no name can be found via the word-level match). Often, since these two matching methods focus on different levels of granularity, their results may not align. Based on the results, you must determine whether each movie name extraction is correct and what the precise movie name for that extracted name should be from the database. 

\vspace{3mm}
\noindent {\color{red}$F^{ref}_{e}$}:  To reflect on this, for each extracted movie, you must respond with three terms separated by \#\#\#\#: (1) the raw movie name mentioned in the dialogue (raw refers to the exact text from the dialogue), (2) the precise movie name selected from fuzzy match or BM25 (reply with a space if the movie name extraction is incorrect or if neither match is precise), and (3) the correct extraction method, choosing from [fuzzy, BM25, none, both]. If the fuzzy match and BM25 results differ but both are probable, select the more probable one based on context as the correct name. List the reflection on each movie name in the exact form of raw \_name\#\#\#\#correct\_name\#\#\#\#method on a new line with no additional terms or sentences. 

\vspace{3mm}
\noindent {\color{red}$s_{t}$}: Here is the user's query: \{\}. 

\vspace{3mm}
\noindent {\color{red}$\mathcal{I}^{char}_{t}, \mathcal{I}^{word}_{t}$}:
Here are extracted movie names, fuzzy matches, and BM25 matches from the movie database in the form of extracted\_name\#\#\#\#fuzzy\_match\#\#\#\#BM25\_match: \{\}.
\end{mdframed}

\vspace{3mm}
\noindent\hrulefill

\vspace{1mm}
\noindent \textbf{Eq. (\ref{eq:ref_tre}): Reflection on the Collaborative Retrieval}

\vspace{3mm}

\begin{mdframed}[backgroundcolor=black!10] 

\noindent {\color{red}$T^{aug}$}: Pretend you are a movie recommender system. I will give you a conversation between a user and you (a recommender system), as well as movies retrieved from the movie database based on the similarity with movies mentioned by the user in the context. You need to judge whether each retrieved movie is a good recommendation based on the context.

\vspace{1mm}
\noindent {\color{red}$F^{aug}$}: You need to reply with the judgment of each movie in a line, in the form of movie\_name\#\#\#\#judgment, where judgment is a binary number 0, 1. Judgment 0 means the movie is a bad recommendation, whereas judgment 1 means the movie is a good recommendation. 

\vspace{1mm}
\noindent {\color{red}$C_{:k-1}$}: Here is the conversation: \{\}. 

\vspace{1mm}
\noindent {\color{red}$\mathcal{I}^{CR}_{k}$}: Here are retrieved movies: \{\}.
\end{mdframed}

\vspace{2mm}
\noindent\hrulefill
\vspace{2mm}

\noindent \textbf{Eq. (\ref{eq:rec}): LLM-based Recommendations}

\vspace{2mm}
\begin{mdframed}[backgroundcolor=black!10] 
\noindent {\color{red}$T^{rec}$}: Pretend you are a movie recommender system. I will give you a conversation between a user and you (a recommender system). Based on the conversation, you need to reply with 20 movie recommendations without extra sentences. 

\vspace{1mm}
\noindent {\color{red}$F^{rec}$}: List the standardized title of each movie on a separate line. 

\vspace{1mm}
\noindent {\color{red}$C_{:k-1}$}:  Here is the conversation: \{\}. 

\vspace{1mm}
\noindent {\color{red}$I^{aug}_{s,k}$}: Based on movies mentioned in the conversation, here are some movies that are usually liked by other users: \{\}. 

\vspace{1mm}
\noindent {\color{blue}\textit{rag} prompt (GPT-4o)}: Use the above information at your discretion (i.e., do not confine your recommendation to the above movies).

\noindent {\color{blue}\textit{rec} prompt (GPT-4)}: Consider using the above movies for recommendations."

\end{mdframed}

\vspace{2mm}
\noindent\hrulefill
\vspace{2mm}

\noindent \textbf{Eq. (\ref{eq:ref_rerank}): Reflect and Rerank}

\vspace{2mm}
\begin{mdframed}[backgroundcolor=black!10] 
\noindent {\color{red}$T^{r\&r}$}:  Pretend you are a movie recommender system. I will give you a conversation between a user and you (a recommender system), as well as some movie candidates from our movie database. You need to rate each retrieved movie as recommendations into five levels based on the conversation: 2 (great), 1 (good), 0 (normal), -1 (not good), -2 (bad). 

\vspace{1mm}
\noindent {\color{red}$F^{r\&r}$}: You need to reply with the rating of each movie in a line, in the form of movie\_name\#\#\#\#rating, where the rating should be an Integer, and 2 means great, 1 means good, 0 means normal, -1 means not good, and -2 means bad. 

\vspace{1mm}
\noindent {\color{red}$C_{:k-1}$}:  Here is the conversation: \{\}. 

\vspace{1mm}
\noindent {\color{red}$\mathcal{I}^{rec}_{k}$}: Here are the movie candidates: \{\}. 
\end{mdframed}

\vspace{1mm}

\section{Experimental Results on GPT-4 Backbone}
\label{sec:gpt4}
In this subsection, we provide the experimental results of \texttt{CRAG} with GPT-4 backbone on the \textbf{Reddit-v2} dataset. Please note that when generating collaborative retrieval augmented recommendations with Eq. (\ref{eq:rec}), we use the \textit{rec} prompt instead of the \textit{rag} prompt.

\subsection{Analysis of the Two-step Reflections}

We first run the experiments of \texttt{CRAG} with GPT-4 backbone under the setting of Section \ref{sec:exp} to explore the effects of the two reflection processes in \texttt{CRAG} on the model performance. The results are summarized in Fig. \ref{fig:wrt_k_4}. From the figure we can find that the three \texttt{CRAG} variants follow the same trend with the case with GPT-4o backbone, where \textbf{\textit{(i)}} naive collaborative retrieval in \texttt{CRAG-nR12} hurts the performance due to the introduction of context-irrelevant collaborative information, \textbf{\textit{(ii)}} after introducing the context-aware reflection step, \texttt{CRAG-nR2} improves the item coverage compared with \texttt{CRAG-nR12} but it still struggles with item rank, and \textbf{\textit{(iii)}} by adding reflect-and-rerank on top of \texttt{CRAG-nR2}, the proposed \texttt{CRAG} leads to the prioritization of more relevant items at the top positions.

\subsection{Comparison with Baselines}

\begin{figure}[t]
\centering
\includegraphics[width=0.45\textwidth]{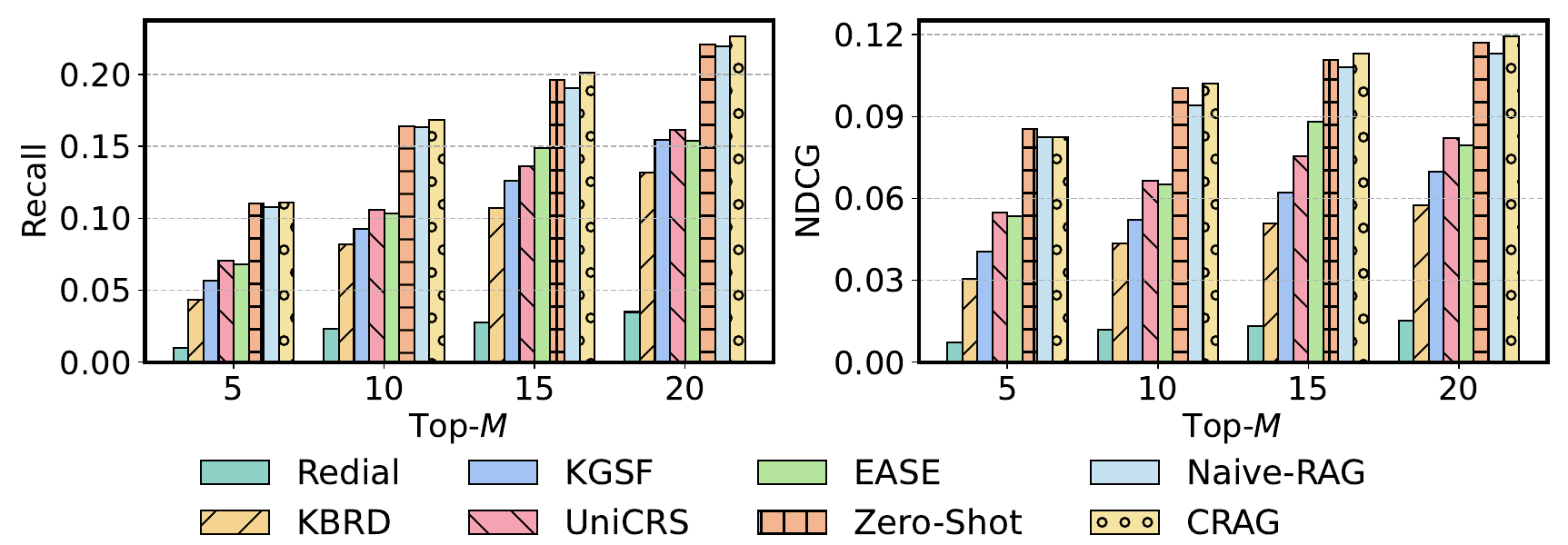}  
\vspace{-1mm} 
\caption{Comparison of CRAG (with GPT-4 backbone) to the baselines on the conversations with no item mentions in the Reddit-v2 dataset.} 
\label{fig:baseline_gpt4}
\end{figure}

We then compare the \texttt{CRAG} with GPT-4 backbone with the baselines introduced in Section \ref{sec:exp2} of the main paper. The results are illustrated in Fig. \ref{fig:baseline_gpt4}, where the relative performance among the methods remains the same as the results of \texttt{CRAG} with GPT-4o backbone illustrated in Fig. \ref{fig:baseline}. In addition, \texttt{CRAG} with GPT-4 backbone also achieves the best performance compared with all the baselines.

\subsection{Evaluation w.r.t. the Recency of Items}

\begin{figure}[t]
\centering
\includegraphics[width=0.45\textwidth]{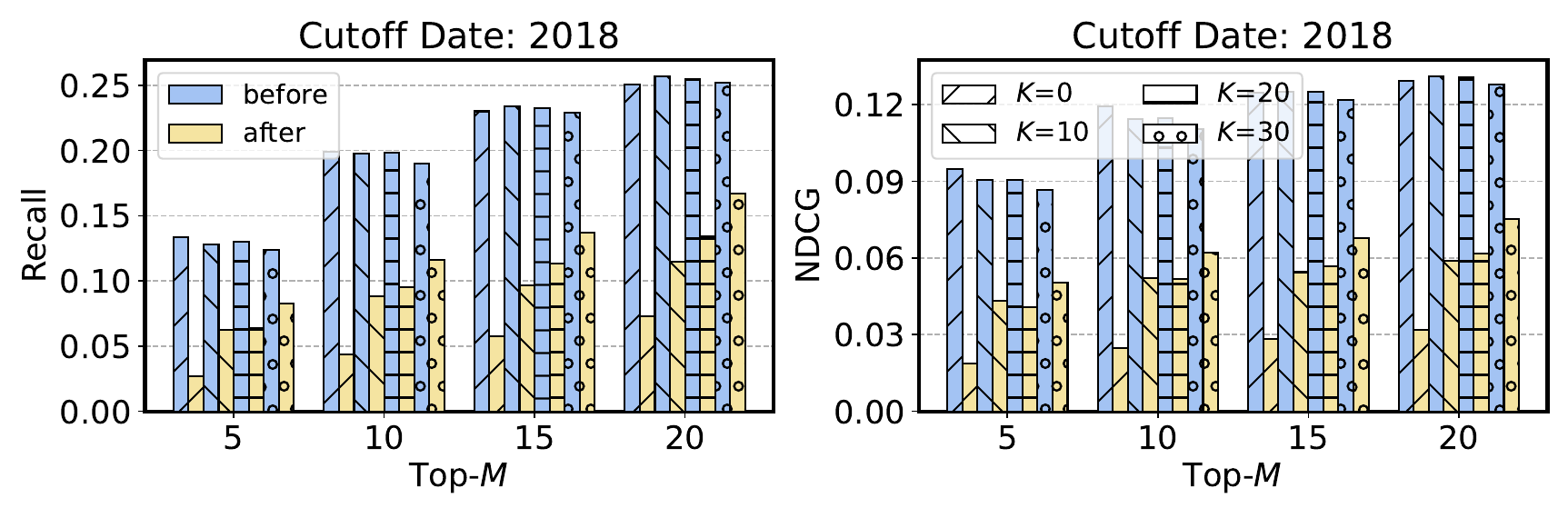}  
\vspace{-1mm}  
\caption{Results of CRAG (with GPT-4 backbone) on the Reddit-v2 dataset where the conversations are separated by the release year of the movies to be recommended.} 
\label{fig:cutoff_gpt4}
\vspace{-3mm}   
\end{figure}

Finally, we evaluate the performance of \texttt{CRAG} with GPT-4 backbone w.r.t. the item recency. Since the cut-off date of GPT-4 is two years before that of GPT-4o, we set the cut-off year (to split the test data) to \texttt{2018}, which is also two years prior to the cut-off year used in Fig. \ref{fig:cutoff} for GPT-4o backbone, and evaluate the model on the  \texttt{before} and \texttt{after} groups with the same setting as Section \ref{sec:abl} in the main paper. The results are summarized in Fig. \ref{fig:cutoff_gpt4}. From Fig. \ref{fig:cutoff_gpt4} we can come to the same conclusion that the improvement for \texttt{CRAG} with GPT-4 backbone over the \texttt{zero-shot LLM} model is largely due to the increased accuracy in recommendations of more recent items.

\begin{table*}[t]
\caption{Comparison between Reddit-v2 and the original Reddit dataset for item extraction. The movie names that the original Reddit dataset extracts incorrectly are marked in {\color{red}\textbf{red}}. The evidence that supports our extraction in the \underline{user query} is highlighted in both \colorbox{red!30}{red} and \colorbox{movie!30}{yellow boxes}, where the \colorbox{red!30}{red boxes} denote the movies that the original Reddit dataset fails to extract.} 
\label{tab:noisy vs clean context}
    \centering
    \begin{tabular}{p{0.05 \textwidth}p{0.35\textwidth}p{0.25\textwidth}p{0.25\textwidth}}
        \textbf{Index} & \textbf{Context} & \textbf{Reddit-v2} & \textbf{Original Reddit}  \\
        \hline 
        \hline
        59 & ...i have watched \colorbox{red!30}{10 things i hate about you} and its my absolute favorite, so im trying to find movies similar to 10 things i hate about you... & 1. 10 Things I Hate About You & \noindent {\color{red}\textbf{NONE}} \\ \\ 
        85 & ...Movies about exploration?. I love \colorbox{red!30}{Master and } \colorbox{red!30}{Commander} and I was thinking about movies about naval exploration...? Thanks' & 1. Master and Commander: The Far Side of the World & \noindent {\color{red}\textbf{NONE}} \\ \\
        155 & ...I have been looking for movies based on small american towns...The only movie that comes to my mind is \colorbox{red!30}{It}... & 1. It & \noindent {\color{red}\textbf{NONE}} \\ \\ 
        156 & Revenge movies?. Looking for something like \colorbox{movie!30}{Kill Bill} or \colorbox{movie!30}{John Wick}. Would be very nice if it's on Netflix or Amazon Prime... & 1. Kill Bill: Vol. 2; 2. John Wick' & \noindent {\color{red}\textbf{1. Revenge;}} {\color{red}\textbf{2. Wild Bill;}} 3. John Wick \\ \\ 
        
        204 & Greatest cast in a movie?. I’d have to say \colorbox{movie!30}{Harlem nights}! Great movie, great cast and funny from start to finish! Eddie Murphy Richard Pryor Red foxx Arsenio Hall Charlie Murphy & 1. Harlem Nights & 1. Harlem Nights; {\color{red}\textbf{2. Red Fox}} \\ \\ 
        219 & Dream films. \colorbox{movie!30}{Inception} is such a great film and I’ve not so much other films attempt a similar premise. So looking for those kinda films where people enter dreams or it has a dream-like state.' & 1. Inception; & \noindent {\color{red}\textbf{1. Dream Kiss;}} 2. Inception \\ \\ 
        
        243 & ...Some examples are: \colorbox{red!30}{Last King of Scotland}, \colorbox{movie!30}{A Bronx Tale}, and \colorbox{movie!30}{Gangs of New York}. I dunno why, but I love these types of films... & 1. The Last King of Scotland; 2. A Bronx Tale; 3. Gangs of New York & 1. A Bronx Tale; 2. Gangs of New York; {\color{red}\textbf{3. NONE}}  \\ \\ 
        
        606 & Need movie like \colorbox{movie!30}{Eyes Wide Shut}. Already watched \colorbox{red!30}{Archive 81} that had masque secret society...Looking for movies about the wealthy elite like Rothchilds. & 1. Eyes Wide Shut; 2. Archive 81 & 1. Eyes Wide Shut; {\color{red}\textbf{2. Archive;}} {\color{red}\textbf{3. Archive;}} {\color{red}\textbf{4. Rothchild}} \\ \\ 

        639 & I am looking for every version of "A Christmas Carol" ever made.. Putting together a bit of a holiday film fest/challenge. I am looking for every version/adaptation of A Christmas Carol that has ever been made, from \colorbox{movie!30}{Scrooged} to \colorbox{movie!30}{Muppets}. & 1. Scrooged; 2. The Muppet Christmas Carol & \noindent {\color{red}\textbf{1. A Christmas Carol;}} 2. Scrooged; {\color{red}\textbf{3. Puppets}} \\ \\ 
        710 & Out of nowhere Children's Horror?. I was just watching \colorbox{movie!30}{The Care Bears Movie (1985)} and there is no way it can't be classified as Children's Horror. Is there any other unexpected horror in Children's IP?...& 1. The Care Bears Movie & 1. The Care Bears Movie; {\color{red}\textbf{2. Children's War}}  \\ \\ 
        \hline
        \hline   
    \end{tabular}
\end{table*}

\begin{table*}[t]
\caption{Comparison between Reddit-v2 and the original Reddit dataset for item extraction. The movie names that the original Reddit dataset extracts incorrectly are marked in {\color{red}\textbf{red}}. The evidence that supports our extraction in the \underline{system response} is highlighted in both \colorbox{red!30}{red} and \colorbox{movie!30}{yellow boxes}, where \colorbox{red!30}{red boxes} denote movies that the original Reddit dataset fails to extract.} 
\label{tab:noisy vs clean response}

    \centering
    \begin{tabular}{p{0.05 \textwidth}p{0.35\textwidth}p{0.25\textwidth}p{0.25\textwidth}}
        \textbf{Index} & \textbf{Response} & \textbf{Reddit-v2} & \textbf{Original Reddit}  \\ \\ 
        \hline 
        \hline
    5 & \colorbox{movie!30}{Mermaids}, \colorbox{movie!30}{Scent of a Woman}, \colorbox{movie!30}{Mickey Blue Eyes}, \colorbox{movie!30}{Mystic Pizza}, and \colorbox{red!30}{Rainy Day in NY} & 1. Mermaids; 2. Scent of a Woman; 3. Mickey Blue Eyes; 4. Mystic Pizza; 5. A Rainy Day in New York & 1. Mermaids; 2. Scent of a Woman; 3. Mickey Blue Eyes; 4. Mystic Pizza; {\color{red}\textbf{5. NONE}} \\ \\ 
    15 & Cocteau’s ‘\colorbox{movie!30}{Orpheus}’ it’s like exactly what you’re looking for You might also like Jarmusch’s ‘\colorbox{red!30}{Paterson}’ and Van Sant’s ‘\colorbox{movie!30}{Drugstore Cowboy}’ and ‘\colorbox{movie!30}{My Own Private Idaho}’ & 1. Orpheus; 2. Paterson; 3. Drugstore Cowboy; 4. My Own Private Idaho & 1. Orpheus; 2. Drugstore Cowboy; 3. My Own Private Idaho \\ \\ 
    61 & \colorbox{red!30}{Man bites dog}, \colorbox{movie!30}{Martin and orloff}, \colorbox{movie!30}{the doom generation} & 1. Man Bites Dog; 2. Martin \& Orloff; 3. The Doom Generation & 1. Martin \& Orloff \\ \\ 
    74 & \colorbox{movie!30}{Baise-moi} \colorbox{movie!30}{Shortbus} \colorbox{red!30}{Nymphomaniac Nymphomaniac 2} & 1. Baise-moi; 2. Shortbus; 3. Nymphomaniac: Vol. I; 4. Nymphomaniac: Vol. II & 1. Baise-moi; 2. Shortbus \\ \\ 
    133 & You listed \colorbox{red!30}{Conan}, are you lumping Red Sonja into the Conan franchise. Just ensuring you haven't missed that one. &  1. Conan; 2. Red Sonja & \noindent {\color{red}\textbf{1. Conman;}} {\color{red}\textbf{2. Conman}}; 3. Red Sonja \\ \\ 
    159 & \colorbox{red!30}{the harder they fall}, it’s on netflix also \colorbox{movie!30}{the crow} & 1. The Harder They Fall; 2. The Crow & 1. The Crow \\ \\ 
    172 & \colorbox{red!30}{The second and third Die Hard movies} all take place within 24 hours as well. & 1. Die Hard 2; 2. Die Hard with a Vengeance & \noindent {\color{red}\textbf{1. Die Hard}} \\ \\ 
    105 & ...It gets mentioned a lot here but **\colorbox{movie!30}{Amelie}** is a movie that always lifts me up. This year I’d also recommend **\colorbox{movie!30}{Everything, Everywhere, All at Once}**'... & 1. Amelie; 2. Everything Everywhere All at Once & 1. Amelie; {\color{red}\textbf{2. Everything;  3. Everywhere;  4. All at Once}} \\ \\ 
    207 & Gotta be \colorbox{red!30}{It's a Mad, Mad, Mad, Mad World}. & 1. It's a Mad Mad Mad Mad World & \noindent {\color{red}\textbf{1. The Longest Day; 2.  The Longest Day}} \\ \\ 

    269 & *\colorbox{movie!30}{North By Northwest}* (1959). A bit like a Bond film before Bond. Hitchcock. Very stylish. Cary Grant and Eva Marie Saint. & 1. North by Northwest & 1. North by Northwest; {\color{red}\textbf{2. Bound; 3. Bound; 4. Bound; 5. Bound}}\\ \\ 
    308 &  \colorbox{movie!30}{Lock, Stock, and Two Smoking Barrels}. \colorbox{movie!30}{In Bruges}. \colorbox{red!30}{And There Were None} (either the 1945 movie or the 2015 mini-series with Charles Dance). & 1. Lock, Stock and Two Smoking Barrels; 2. In Bruges; 3. And Then There Were None & \noindent {\color{red}\textbf{1. Lock; 2. Stuck; 3. Lock;}} 4. In Bruges \\ \\ 
    \hline 
    \hline
    \end{tabular}
\end{table*}

\begin{table*}[t]
    \caption{Examples of movies identified with positive attitude in the established Reddit-v2 dataset. The movie names are marked with \colorbox{green!30}{green boxes} in the user query or the system response in the context column.}  
    \label{tab: positive}
    \centering
    \begin{tabular}{p{0.05 \textwidth}p{0.5\textwidth}p{0.3\textwidth}}
        \textbf{Index} & \textbf{Context} & \textbf{Extracted movie names}  \\
        \hline 
        \hline
        555 & ...Here is a list of movies that absolutely ruined me for weeks, some still haunt me with late night horror of being someone’s victim simply because “You were home” \colorbox{green!30}{1. The Strangers; 2. Eden Lake; 3. Funny;} \colorbox{green!30}{Games; 4. Zodiac; 5. The Last House on the Left}...  & 1. The Strangers; 2. Eden Lake; 3. Funny Games; 4. Zodiac; 5. The Last House on the Left \\ \\ 
        500 & ...I feel like since the covid lockdown I've seen like every scifi action movie of this millenium...Things in the vein of the more modern \colorbox{green!30}{AvP} movies, \colorbox{green!30}{Battle of LA}, the Frank Grillo and his son fighting aliens series that I'm blanking on the name of, \colorbox{green!30}{Pacific Rim} franchise, etc... & 1. Alien vs. Predator; 2. Battle Los Angeles; 3. Pacific Rim \\ \\ 
        519 & Best Foreign Movies?. I recently watched \colorbox{green!30}{Troll} and \colorbox{green!30}{Pans Labyrinth}. I wasn't always fond of movies with subtitles but I really enjoy them now. What are some good Sci-fi/Fantasy foreign films? & 1. Troll; 2. Pan's Labyrinth \\ \\ 
        544 & Most Disturbing WW2 movies. Alright guys I saw \colorbox{green!30}{all quiet on the western front} the other night and I really enjoyed it. I’m looking for the most bloodiest war movie you can recommend me. Preferably WW2 & 1. All Quiet on the Western Front \\ \\ 
        554 & Time loop movies. There are several great time loop movies out there, and some of my favorites include: \colorbox{green!30}{Groundhog Day} - In this classic comedy, a weatherman finds himself reliving ... to become a better person. \colorbox{green!30}{Happy Death Day} - A college student must relive the day of her murder over and over again until she figures out who the killer is. \colorbox{green!30}{Edge of Tomorrow} -... & 1. Groundhog Day; 2. Happy Death Day; 3. Edge of Tomorrow \\ \\ 
        576 & I'm looking for movies with a global threat.. Specifically a movie where a bunch of organizations ... come together and work to understand, fight, and hopefully defeat it. The only example I can think of right now is "\colorbox{green!30}{Contagion}". I greatly appreciate any and all suggestions :) Thank you! & 1. Contagion \\ \\ 
        701 & ...I'm looking for something more where the movie's plot would go on and just display that the male's love interest or actress just happens to be older than him and that's it. An example of this is \colorbox{green!30}{Water for Elephants} where Reese Witherspoon is ten years older than Robert Pattinson, but the film still focuses on the circus storyline... & 1. Water for Elephants  \\ \\ 
        879 & the funniest non mainstream comedy.. I'm looking for a good comedy that I haven't seen before. I love comedy's like \colorbox{green!30}{odd couple 2, palm springs, the wrong missy, vacation (2015),nothing } \colorbox{green!30}{to lose}. Movies like the hang over, super bad are just so stale and overrated. Any suggestions please? I need a good laugh tonight. & 1. The Odd Couple II; 2. Palm Springs; 3. The Wrong Missy; 4. Vacation; 5. Nothing to Lose\\ \\ 

        938 & ...Movies like \colorbox{green!30}{Mean Girls} and \colorbox{green!30}{Freaky Friday}?. I really like these two movies. not particularly because of Lindsay btw although I liked her on these movies. are there like "go to movies" that are similar to these?... & 1. Mean Girls; 2. Freaky Friday \\ \\ 
        \hline
        \hline
        
    \end{tabular}

\end{table*}

\begin{table*}[t]
    \caption{Examples of movies identified with neutral attitude in the established Reddit-v2 dataset. The movie names are marked with \colorbox{movie!30}{yellow boxes} in the user query or the system response in the context column.}  
    \label{tab: neutral}
    \centering
    \begin{tabular}{p{0.05 \textwidth}p{0.5\textwidth}p{0.3\textwidth}}
        \textbf{Index} & \textbf{Context} & \textbf{Extracted movie names}  \\
        \hline 
        \hline
        607 & What would you consider "must-see" movies?. I\'m sorry if this has been asked a million and one times, I\'m new here...Every time I look at lists of favorite movies, they always seem to be the same things, \colorbox{movie!30}{Citizen Kane, Shawshank, Godfather, Casablanca}, etc. And no hate to those movies!! But they\'re classics for a reason, I've already seen them and want something new!... & 1. Citizen Kane; 2. The Shawshank Redemption; 3. The Godfather; 4. Casablanca \\ \\ 
        683 & Movies about guns.. I’m seeking films about guns or involving lots of gun action. For example: \colorbox{movie!30}{Lord Of War Gun Crazy Hardcore Henry} I am going to just fill the rest here for the mandatory text limit because I have nothing else to say. Please comment below. & 1. Lord of War; 2. Gun Crazy; 3. Hardcore Henry \\ \\ 
        1035 & Akira (1988) Is an amazing film. Akira (1988), which I saw for the first time last night, completely floored me. I can't believe I haven't seen the film sooner after having it on my to-do list for so long. I'm not a huge anime fan \colorbox{movie!30}{Spirited Away} and \colorbox{movie!30}{Pokémon} are about the extent of my knowledge), but I think anyone would like this film... & 1. Spirited Away; 2. Pokémon \\ \\
        1474 & ... I would like to see some movies where the main character or an important character is red haired, i don't mind if it's natural or not. Last movie i saw was  \colorbox{movie!30}{Perfume: The Story of a Murderer}  and i was wondering why red haired/gingers women are so rare in movies. I would appreciate even movies where the girl is not the protagonist, tho keep in mind she should be on the screen more then 1 scene. Any type of movie is welcomed. Thank you in advance. & 1. Perfume: The Story of a Murderer \\ \\ 
        1395 & My wife is currently getting a procedure done that will leave her face appearing severely burned for several days. Other than Nicolas Cage’s \colorbox{movie!30}{Face/Off}, what movies should I queue for our marathon while she recovers?... & 1. Face/Off \\ \\
        1673 & Best of the Middle East. I had a chance to watch...I would love to see more great Egyptian/Middle Eastern/Arabic/North African films. Other than the Iranian **\colorbox{movie!30}{A Girl Walks Home Alone At Night}** I haven't really seen much of anything from the region. Any suggestions on where to start?" & 1. A Girl Walks Home Alone at Night \\ \\ 
        1690 & ...I’m asking this because I’m watching \colorbox{movie!30}{Thor: Love and Thunder} for the first time and while it’s not bad, it feels more like background noise or standard popcorn fare.  It’s fine and all but it got me thinking, what are some movies where my attention will be absolutely grabbed?  Where pulling out my phone even to look at it for a second would be unwanted?  & 1. Thor: Love and Thunder \\ \\
        2903 & Sequels which pick up immediately from the original. What movies pick up exactly from where their originals leave off? I don\'t mean "a short while later" like \colorbox{movie!30}{Star Wars: A New Hope to The Empire Strikes Back}, but straight shots with continuity... & 1. Star Wars: Episode IV - A New Hope; 2. Star Wars: Episode V - The Empire Strikes Back \\ \\ 
        
        \hline
        \hline
        
    \end{tabular}

\end{table*}

\begin{table*}[t]
    \caption{Examples of movies identified with negative attitude in the established Reddit-v2 dataset. The movie names are marked with \colorbox{red!30}{red boxes} in the user query or system response in the context column.}  
    \label{tab: negative}
    \centering
    \begin{tabular}{p{0.05 \textwidth}p{0.5\textwidth}p{0.3\textwidth}}
        \textbf{Index} & \textbf{Context} & \textbf{Extracted movie names}  \\
        \hline 
        \hline

        590 & And please don\'t give me the shallow happy-go-lucky "\colorbox{red!30}{Fundamentals of Caring}" type of shit. I need deep, relatable emotions and metaphysical devastation. If I don\'t bawl at the screen questioning every God\'s existance towards the end, it was not worth it. & 1. The Fundamentals of Caring \\ \\ 
        701 & ...I am looking for a movie where a younger man and an older woman develop a romantic relationship...but it wouldn't be anything like \colorbox{red!30}{The Graduate}, or \colorbox{red!30}{The Piano Teacher} where their age gap is treated as taboo and is the centered plot... & 1. The Graduate; 2. The Piano Teacher \\ \\
        879 & ...Movies like the \colorbox{red!30}{hang over}, \colorbox{red!30}{super bad} are just so stale and overrated. Any suggestions please? I need a good laugh tonight. & 1. The Hangover; 2. Superbad \\ \\ 
        1061 & What’s the best (bad) Christmas movie.. Bad Christmas movies are a guilty pleasure of mine...What are your favorite bad movies?  Major studio release, or made for tv trash, I don’t care. Just tell me the movie, who’s in it, and a simple plot, if I haven’t seen it, I’ll go find it. No “good” movies though.  Don’t recommend \colorbox{red!30}{White Christmas} or “\colorbox{red!30}{it’s a wonderful life}” not only do we all know them, but they are iconic... & 1. White Christmas; 2. It's a Wonderful Life \\ \\ 
        1092 & Can you suggest some Netflix series that is for people who are really alone... For eg., I was watching the new \colorbox{red!30}{Wednesday} series and hoping that I could relate to Wednesday Addams, only to realize that it is just another teen drama where supposedly lonely and evil Wednesday Addams has multiple love interests and saves... & 1. Wednesday \\ \\
        1131 & actually scary zombie/vampire movies?. I watched \colorbox{red!30}{28 Days Later} which I've heard is scary but I found it rather boring. I also watched \colorbox{red!30}{Braindead} but it wasn't scary, just gross. As for vampire movies, I love them but I've never seen any that is actually scary to me. What do you think?... & 1. 28 Days Later; 2. BrainDead \\ \\ 
        1160 & Intense romance with a happy and fulfilling ending.. I just watched \colorbox{red!30}{King Kong (2006)} and now I feel hollow inside. So sad. It's like an intense romance with a tragic ending so now I need an intense romance with an extremely fulfilling ending where the two lovers go through intense hardships...& 1. King Kong \\ \\ 
        1335 & I'm looking for quality story sci-fi / fantasy from 2010-20s... What I mean is, i tried watch "\colorbox{red!30}{Life}" to find an fascinatic newer sci-fi, ended up being close to brutal and grotesque. I tried watching \colorbox{red!30}{4400} series, ended up being not that much about sci-fi but about trans/lesbian activism, teenage romance dramas, anti-christian activism... & 1. Life; 2. 4400 \\ \\ 
        1353 & Movies with interracial relationships, that aren't strictly ABOUT that?. So not stuff like \colorbox{red!30}{Jungle Fever}, \colorbox{red!30}{Get Out}, etc. Films that could be in any genre, not just romance. The films can be I guess from any year, ideally in colour, but lean towards the '80s...& 1. Jungle Fever; 2. Get Out \\ \\
        \hline
        \hline
    \end{tabular}
\end{table*}

\end{document}